\documentclass[useAMS,usenatbib]{mnras}
\pdfoutput=1 
\usepackage{amsmath,amstext}
\usepackage[T1]{fontenc}
\usepackage{graphicx}
\usepackage[dvipsnames]{xcolor}
\usepackage[hang]{subfigure}
\usepackage{appendix}
\usepackage{array,multirow}
\usepackage{amssymb}
\usepackage{hyperref}
\usepackage{lmodern}
\usepackage{booktabs}
\usepackage{multirow}
\usepackage{comment}
\usepackage{pifont}
\usepackage{tabularx}
\usepackage{makecell}
\usepackage{titlesec}
\usepackage[normalem]{ulem}
\newcommand{\txn}[1]{\textnormal{#1}}

\newcommand{\beagle}{\textsc{beagle}}
\newcommand{\beagleagn}{\textsc{beagle-agn}}
\newcommand{\pypbeagle}{\textsc{pyp-beagle}}
\newcommand{\cloudy}{\textsc{cloudy}}

\newcommand{\M}{\hbox{$\txn{M}$}}

\newcommand{\MstarInLog}{\hbox{$\txn{M}$}}

\newcommand{\Msun}{\hbox{$\M_{\sun}$}}
\def\Msun{\hbox{$\rm\thinspace M_{\odot}$}}

\newcommand{\Mtot}{\hbox{$\M_\txn{tot}$}}
\newcommand{\MtotInLog}{\hbox{$\txn{M}_\txn{tot}$}}
\newcommand{\yr}{\hbox{$\txn{yr}$}}

\renewcommand{\t}{\hbox{$t$}}

\newcommand{\tausfr}{\hbox{$\tau_\textsc{sfr}$}}

\newcommand{\sfr}{\hbox{${\Psi}$}}
\newcommand{\sfrInLog}{\hbox{${\psi}$}}

\newcommand{\Z}{\hbox{$\txn{Z}$}}

\newcommand{\Zhii}{\hbox{$\Z_\txn{gas}^\textsc{hii}$}}

\newcommand{\Zsun}{\hbox{$\Z_{\odot}$}}

\newcommand{\HII}{\mbox{H\,{\sc ii}}}

\newcommand{\nH}{\hbox{$n_\textsc{h}$}}

\newcommand{\logUs}{\hbox{$\log U_\textsc{s}$}}

\newcommand{\xid}{\hbox{$\xi_\txn{d}$}}

\newcommand{\tauV}{\hbox{$\hat{\tau}_\textsc{v}$}}
\newcommand{\tauVnlr}{\hbox{$\hat{\tau}_\textsc{v}^{NLR}$}}
\newcommand{\mud}{\hbox{$\mu$}}

\newcommand{\Lacc}{\hbox{$L_\txn{acc}$}}
\newcommand{\PLalpha}{\hbox{$\alpha_\textsc{pl}$}}

\newcommand{\logUsAGN}{\hbox{$\log U_\textsc{s}^\textsc{nlr}$}}

\newcommand{\UsAGN}{\hbox{$U_\txn{s}^\textsc{nlr}$}}
\newcommand{\xidAGN}{\hbox{$\xi_\txn{d}^\textsc{nlr}$}}
\newcommand{\ZAGN}{\hbox{$\Z_\txn{gas}^\textsc{nlr}$}}
\newcommand{\nHAGN}{\hbox{$n_\textsc{h}^\textsc{nlr}$}}

\newcommand{\oii}{\relax \ifmmode {\mbox O\,{\scshape ii}}\else O\,{\scshape ii}\fi}
\newcommand{\oiii}{\relax \ifmmode {\mbox O\,{\scshape iii}}\else O\,{\scshape iii}\fi}
\newcommand{\OIIIauroral}{\hbox{[O{\sc iii}]$\lambda4363$}}
\newcommand{\Halpha}{\hbox{H$\alpha$}}
\newcommand{\Hbeta}{\hbox{H$\beta$}}
\newcommand{\Hgamma}{\hbox{H$\gamma$}}
\newcommand{\Hdelta}{\hbox{H$\delta$}}
\newcommand{\Hepsilon}{\hbox{H$\epsilon$}}
\newcommand{\OI}{\hbox{[O\,{\sc i}]$\lambda6300$}}

\newcommand{\OII}{\hbox{[O\,{\sc ii}]$\lambda3726,\lambda3729$}}

\newcommand{\OIIIfi}{\hbox{[O\,{\sc iii}]$\lambda5007$}}
\newcommand{\OIIIft}{\hbox{[O\,{\sc iii}]$\lambda4363$}}
\newcommand{\OIIIfn}{\hbox{[O\,{\sc iii}]$\lambda4959$}}

\newcommand{\CIV}{\hbox{[C\,{\sc iv}]$\lambda1549$}}

\newcommand{\NIV}{\hbox{[N\,{\sc iv}]$\lambda1486$}}

\newcommand{\NeIII}{\hbox{[Ne\,{\sc iii}]$\lambda3869$}}
\newcommand{\NeIV}{\hbox{[Ne\,{\sc iv}]$\lambda2426$}}
\newcommand{\teNeIII}{\hbox{$t_e$(Ne\sc iii)}}

\newcommand{\newchange}[1]{#1} 
\newcommand{\newchangemath}[1]{#1} 

\newcolumntype{L}[1]{>{\raggedright\let\newline\\\arraybackslash\hspace{0pt}}m{#1}}
\newcolumntype{C}[1]{>{\centering\let\newline\\\arraybackslash\hspace{0pt}}m{#1}}
\newcolumntype{R}[1]{>{\raggedleft\let\newline\\\arraybackslash\hspace{0pt}}m{#1}}

\newcommand{\ticks}[1]{%
      \reflectbox{'}#1'}

\title[\textsc{beagle-agn SMACS S06355}]{Characterising the z $\approx$ 7.66 Type-II AGN candidate SMACS S06355 using BEAGLE-AGN and JWST NIRSpec/NIRCam}
\author[Silcock et al. ]{M. S. Silcock$^{1}$\thanks{Email: m.s.silcock@herts.ac.uk},
E. Curtis-Lake$^{1}$, D. J. B. Smith$^{1}$, I. E. B. Wallace$^{2}$,  A. Vidal-Garc\'ia$^{3}$, A. Plat$^{4,5}$, \newauthor M. Hirschmann$^{5,6}$, A. Feltre$^{7}$, J. Chevallard$^{2}$, S. Charlot$^{8}$, S. Carniani$^{9}$ and A. J. Bunker$^{2}$\\\\
$^{1}$ Centre for Astrophysics Research, Department of Physics, Astronomy and Mathematics, University of Hertfordshire, Hatfield AL10 \\9AB, UK \\
$^{2}$ Sub-department of Astrophysics, Department of Physics, University of Oxford, Denys Wilkinson Building, Keble Road,
Oxford OX1 \\3RH, UK \\
$^{3}$ Observatorio Astronómico Nacional, C/ Alfonso XII 3, 28014 Madrid, Spain\\
$^{4}$ Steward Observatory, 933 N. Cherry Avenue, University of Arizona, Tucson, AZ 85721, USA \\
$^{5}$ Institute of Physics, GalSpec Laboratory, Ecole Polytechnique Federale de Lausanne, Observatoire de Sauverny, Chemin Pegasi 51, \\ 1290 Versoix, Switzerland \\
$^{6}$ INAF, Osservatorio Astronomico di Trieste, Via G. B. Tiepolo 11, 34131 Trieste, Italy \\
$^{7}$ INAF-Osservatorio Astrofisico di Arcetri, Largo E. Fermi 5, I-50125, Firenze, Italy\\
$^{8}$ Sorbonne Universit\'{e}, CNRS, UMR7095, Institut d'Astrophysique de Paris, F-75014, Paris, France\\
$^{9}$ Scuola Normale Superiore, Piazza dei Cavalieri 7, I-56126 Pisa, Italy}

\begin{document}

\date{}

\pagerange{\pageref{firstpage}--\pageref{lastpage}} \pubyear{2025}

\maketitle

\label{firstpage}

\begin{abstract}
\noindent 
The presence of Active Galactic Nuclei (AGN) in low mass ($\mathrm{M_{\star}\lesssim10^{9}}\,\mathrm{M_{\odot}}$) galaxies at high redshift has been established, and it is important to characterise these objects and the impact of their feedback on the host galaxies. In this paper we apply the Spectral Energy Distribution (SED) fitting code \beagleagn\ to SMACS S06355, a \textit{z}$\approx$7.66 Type-II AGN candidate from the JWST NIRSpec Early Release Observations. This object's spectrum includes a detection of the \NeIV\ line, indicating an obscured AGN due to its high ionization potential energy ($\sim$63eV). 
We use \beagleagn\ to simultaneously model the Narrow Line Region (NLR) AGN and star-forming galaxy contributions to the observed line fluxes and photometry. 
Having a high-ionization emission line allows the contribution of the NLR to the remaining lines to be probabilistically disentangled. The \HII\ region metallicity is derived to be 12+log(O/H)$^{\mathrm{HII}}$ = $\newchangemath{7.82^{+0.18}_{-0.19}}$. Assuming that the Neon-to-Oxygen abundance is similar to solar we derive a high NLR metallicity of 12+log(O/H)$^{\mathrm{NLR}}$ = $\newchangemath{8.86^{+0.14}_{-0.16}}$, with the 2$\sigma$ lower-limit extending to 12+log(O/H)$^{\mathrm{NLR}}\sim$\newchange{8.54}, showing the derivation is uncertain.  We discuss this result with respect to non-solar Neon abundances that might boost the inferred NLR metallicity. The NLR metallicity places SMACS S06355 in a comparable region of the mass-metallicity plane to intermediate (1.5$\lesssim$\textit{z}$\lesssim$3.0) redshift obscured AGN. Our derived accretion disc luminosity, log($\mathrm{\Lacc/erg s^{-1}}$)=$45.19^{+0.12}_{-0.11}$, is moderately high yet still uncertain. We highlight that deviations between bolometric luminosity calibrations and model grid tracks become enhanced at low metallicities. 
\end{abstract}

\begin{keywords}
galaxies: active, galaxies: evolution, galaxies: high-redshift, galaxies: nuclei, galaxies: Seyfert, galaxies: general

\end{keywords}

\newpage

\section{Introduction}
\label{section:introduction}

The first years of \textit{JWST} have delivered many active galactic nucleus (AGN) candidates in the early (\textit{z} $\gtrsim$ 5) Universe \citep{Brinchmann2023,Larson2023, Kocevski2023, Maiolino2023b}. For the first time, we are able to characterise black hole demographics and growth at early times with unprecedented detail, which helps with understanding possible seeding mechanisms in the early Universe \citep{Volonteri2021}. 

Most candidates are Type-I AGN, identified from their broad permitted emission lines due to a direct sightline to the broad line region (BLR) surrounding a super-massive black hole \citep{Kocevski2023, Maiolino2023b}. \newchange{Such broad lines typically exhibit velocity dispersions of approximately $2000$ km s$^{-1}$, and historically lines were considered narrow at $\lesssim1000\ \mathrm{km s}^{-1}$ \citep{Daltabuit&Cox1972}. In practice a Type-I AGN might be identified if the permitted lines are broader than the measured forbidden lines, which would not be the case if observing an AGN for which the BLR is obscured (Type-II AGN), and spectroscopically identifying these Type-II AGN} is proving more challenging \citep{Scholtz2023}. 
One main reason for this is that at lower metallicities, standard rest-frame optical emission-line diagnostics \citep[e.g. the much used BPT/VO87 line ratio diagnostics;][]{Baldwin1981,Veilleux1987} fail to separate AGN and star-forming galaxies \citep{Groves2004, Hirschmann2019, Hirschmann2023, Ubler2023}. \newchange{For example, \citet[fig. 2]{Feltre2016} showed that at low AGN metallicity ($\mathrm{Z_{AGN} \lesssim 0.5\ \Zsun}$), AGN model coverage typically falls below the \cite{Kauffmann2003} SF-Seyfert demarcation line \citep[see also][]{Zhu2023, Dors2024}}. Rest-frame UV emission line diagnostics have been proposed to overcome these problems \citep{Feltre2016,Hirschmann2019}, though they are found to provide different candidates depending on the criteria imposed \citep{Scholtz2023}, and a top-heavy initial mass function (IMF) or Wolf-Rayet (WR) stars may explain strong UV line emission in some cases \citep{Cameron2024, Senchyna2024}. Robust candidates often show high-ionization lines that cannot be explained by standard models of star-formation alone (for example, see fig. 1 of \citealt{Feltre2016}; see also candidates within \citealt{Scholtz2023}).

This paper focuses on one particular object identified from the SMACS J0723.3-7327 (SMACS 7327) cluster field (\citealt{Pontoppidan2022, Carnall2023}), one of the targets from the Early Release Observations (ERO). In the ERO NIRSpec Multi-Shutter Array (MSA) spectroscopy \cite{Brinchmann2023} identified a high-ionization line, \NeIV, in object SMACS S06355\footnote{Here, \ticks{(0)6355} refers to the object's ID, where the \ticks{S} refers to it belonging to the SMACS J0723.3-7327 cluster \citep[see nomenclature in][for example]{Katz2023}} with coordinates RA: 110.84452 and Dec: $-73.43508$ \citep[][table 2]{Carnall2023}. \newchange{The presence of this line requires photons of energy $\geq63.45$ eV (the ionization potential of Ne$^{2+}$) to ensure the presence of triply ionized Neon.  These energies are beyond those} produced by stellar populations characterised by low metallicities and a high upper mass limit cutoff to the IMF of 300\Msun \citep{Lecroq24}.
The object was therefore identified as a probable AGN \citep{Brinchmann2023} at $z=7.6643\pm0.0010$ within weeks of the first data to be released for the telescope. This object is one of three high redshift galaxies from the same dataset to exhibit the \OIIIauroral\ line in the early Universe, an auroral line that, in concert with the brighter \OIIIfi\ line, has been used to estimate the electron temperature and provide robust metallicity estimates. \newchange{\cite{Curti2023a,Trump2023,Schaerer2022} provided direct metallicities for this object, but had not identified the \NeIV\ line and therefore } assumed the observed rest-frame optical emission lines (\OII, \Hepsilon, \Hgamma, \Hbeta, \OIIIft, \OIIIfn\ and \OIIIfi) were powered by star formation alone. 

Here we use a new tool, \beagleagn\ \citep{Vidal-Garcia2024}, a \beagle\ \citep[BayEsian Analysis of GaLaxy sEds,][]{Chevallard2016} extension that simultaneously models stellar emission, dust attenuation, nebular emission from \HII\ regions surrounding young stars as well as the narrow-line emission from Type-II AGN. \newchange{The goals of this study are to fit both emission lines and photometric filter fluxes of SMACS S06355 with }\beagleagn\ \newchange{in order to 
investigate the impact of including an AGN component on the derived physical properties of this object (without assuming that the emission lines are solely powered by star formation or an AGN).} 

Section \ref{section:datameasurementsfitting} describes the data and reduction used in the fitting, along with details of the \NeIV\ flux extraction we performed and the fitting process with \beagleagn. Results from the fitting process are presented in Section \ref{section:results}, where we also briefly discuss the version of the results using an IMF upper mass limit of 300$\Msun$ (as opposed to 100$\Msun$ in the fiducial). We discuss the fit we obtain and derived parameters of SMACS S06355 in Section \ref{section:discussion}, which includes discussions on: comparisons to previous studies of this galaxy; alternative ionising sources of \NeIV; the bolometric luminosity; the importance of neon abundances in modelling; and caveats of this work's approach. This work assumes a standard $\Lambda\mathrm{CDM}$ cosmology with $\mathrm{H_{0} = 67.74 km s^{-1} Mpc^{-1}}$, $\mathrm{\Omega_{m} = 0.3089}$ and $\Omega_{\lambda} = 0.6911$ \citep{Plank2016}.

\section{Data Treatment}\label{section:datameasurementsfitting}

\subsection{Data}
\label{subsec:data}

\begin{figure}
\includegraphics[trim={0.26cm 0.30cm 0.0cm 0.25},clip,width=0.5\textwidth]{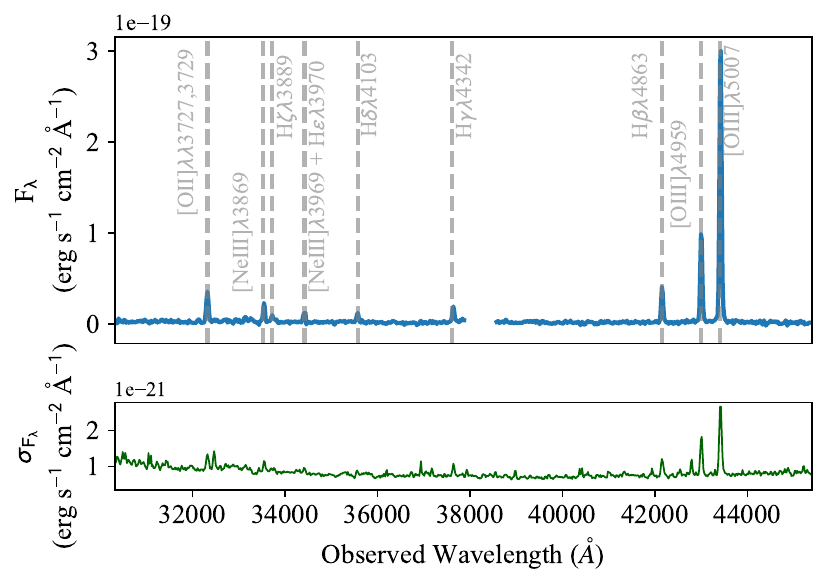}
\caption{The observed G395M/F290LP 1D spectrum of SMACS S06355, shown as a solid blue line in the upper panel. The gray dashed vertical lines and their corresponding rest-frame wavelength labels highlight notable emission lines relevant to this work. The lower panel shows the corresponding error spectrum in solid green line.}
\label{6355spectrum}
\end{figure}

We make use of NIRSpec \citep{Jakobsen22} MSA mode spectroscopy \citep{Ferruit22} and NIRCam \citep{Rieke22} imaging taken of the SMACS 7327 cluster field as part of the Early Release Observations (PI: Klaus Pontoppidan, program-ID: 2736) to perform spectro-photometric fits. Spectra were taken with two grating/filter configurations, G395M/F290LP and G235M/F170LP, covering the wavelength range 1.66$\mu$m to 5.16$\mu$m at a spectral resolution of R$\sim$1000. For visual reference, we include the G395M/F290LP 1D spectrum in \newchange{Figure} \ref{6355spectrum}. In this paper, we focus on the object with ID 6355, which has total exposure time 8840 seconds in each configuration. The measured emission line fluxes from the spectra provide constraints on the ionized gas in the galaxy.  To gain constraints on the stellar component (stellar mass, dust attenuation experienced by stars, etc) we need strong constraints on the stellar continuum and relative strengths of the emission lines (equivalent widths).  In this case the continuum level has a very low S/N in the spectra (\newchange{average signal-to-noise across 5050\AA\ - 5400\AA\ range is S/N = 1.58)}, and so we resort to photometry to provide the continuum level.  The photometric data thus provide constraints on the stellar continuum as well as the emission lines in some of the filters. We used photometry measured from the following NIRCam filters: F090W, F150W, F200W, F277W F356W and F444W.  All line and filter fluxes are uncorrected for magnification by the central cluster.  We apply the appropriate magnification corrections to derived quantities after fitting with \beagleagn\footnote{The emission line fluxes from \cite{Curti2023a} were also non-magnification corrected at the instance of fitting. As part of our procedure we reverse the point source corrections attributed to these fluxes and so to maintain consistency, we also keep emission line fluxes uncorrected for magnification here.}. 

The individual micro-shutters of NIRSpec are small, with an illuminated area of $0.20^{\prime\prime} \times 0.46^{\prime\prime}$, and so an appreciable fraction of the flux will fall outside the slit even for a point source, particularly at longer wavelengths where the point spread function (PSF) is larger. We use the measured line fluxes presented in table 1 of \cite{Curti2023a}, with the exception of \NeIV, which we describe in the following section, and \NeIII\ (see Section~\ref{subsec:NeIII}). These fluxes assume a point like target morphology (at the nominal target coordinates, offset from the centre of the micro-shutter) whereas SMACS S06355 is an extended source \citep[see][fig. 2]{Tacchella2023}. Further, the photometry we use describes the total flux in each filter. Therefore, in order to fit both the emission line fluxes and photometry simultaneously, we needed to
apply an aperture correction to the line fluxes from \cite{Curti2023a}. Applying corrections accounting for these differences allows the use of fluxes which are more consistent with the photometry. The following outlines the overall procedure, and the final emission line fluxes (in addition to the photometric filter fluxes) used in our fitting are given in Table \ref{tab:datatable}.

\begin{table}
    \centering
    \caption{Table compiling data used in this work's \beagleagn\ fit to object SMACS S06355. Both spectroscopic and photometric data were used in the fitting, specifically emission line fluxes and filter fluxes. The emission line fluxes presented in the \ticks{Total Flux} column are those corrected to their total equivalents, as expanded in Section \ref{subsec:data}; the emission line and filter fluxes in this column are in units of $\mathrm{10^{-18}\ erg \, s^{-1} cm^{-2}}$ and $\mathrm{nJy}$, respectively. The \ticks{Total/Original} column provides the ratios of the total equivalent fluxes to the point-source corrected versions originally in \protect\cite{Curti2023a}. For \NeIV, we calculate the \ticks{original} flux value from a procedure validated to be consistent with that of \protect\cite{Curti2023a}, as outlined in Section \ref{subsec:fluxextraction}.}
    \begin{tabular}{ccc}
        \toprule
        Data & Total Flux & Total / Original \\
        
        \hline 
        
        [Ne\textsc{iv}] $\lambda$2426 & 0.65 $\pm$ 0.06 & 1.48 $\pm$ \newchange{0.20} \\ 
        
        [O\textsc{ii}] $\lambda$3726,29 & 3.26 $\pm$ 0.42 & 1.72 $\pm$ \newchange{0.29} \\
        \newchange{[Ne}\textsc{iii}\newchange{] $\lambda$3869} & $\newchangemath{1.63 \pm 0.13}$ & $\newchangemath{1.72 \pm 0.19}$\\
        
        H$\delta$ $\lambda$4103 & 0.94 $\pm$ 0.09 & 1.71 $\pm$ \newchange{0.21} \\
        
        H$\gamma$ $\lambda$4342 & 1.65 $\pm$ 0.09 & 1.70 $\pm$ \newchange{0.13} \\
        
        [O\textsc{iii}] $\lambda$4363 & 0.35 $\pm$ 0.08 & 1.70 $\pm$ \newchange{0.50} \\
        
        H$\beta$ $\lambda$4863 & 4.23 $\pm$ 0.18 & 2.00 $\pm$ \newchange{0.10} \\
        
        [O\textsc{iii}] $\lambda$4959 & 11.19 $\pm$ 0.35 & 1.99 $\pm$ \newchange{0.09} \\
        
        [O\textsc{iii}] $\lambda$5007 & 34.77 $\pm$ 1.01 & 1.99 $\pm$ \newchange{0.09} \\
        
        F090W & -7.30 $\pm$ 2.02 & $-$ \\ 
        
        F150W & 128.78 $\pm$ 2.18 & $-$ \\
        
        F200W & 134.19 $\pm$ 1.89 & $-$ \\
        
        F277W & 172.84 $\pm$ 1.86 & $-$ \\
        
        F356W & 238.51 $\pm$ 1.91 & $-$ \\
        
        F444W & 634.97 $\pm$ 3.28 & $-$ \\ \bottomrule
    \end{tabular}
    \label{tab:datatable}
\end{table}

The line fluxes reported in \citet{Curti2023a} are measured from spectra processed by the NIRSpec/GTO pipeline (Carniani, in prep.). The pipeline starts with level 2 data products (count rate maps) and initially performs the pixel-to-pixel  background subtraction by using the three nodding exposures. It then identifies and extracts the spectrum of the selected target. The 2D spectrum is then corrected for flat-field, wavelength calibration, flux calibration, and slit-loss correction assuming a point-like target morphology. Finally, the pipeline rectifies and interpolates the 2D map on to a regular wavelength grid and a 1D spectrum is extracted from each exposure. The final products are the combination of all 1D and 2D spectra of the selected target.

To account for the extended morphology of SMACS S06355, we apply a correction to the emission line fluxes using photometry from the NIRCam images. We first reverse the point source correction by dividing the fluxes by their corresponding path loss correction factor \citep[priv. comm]{Curti2023a}, to recover the flux which actually fell within the micro-shutter aperture. Total photometric fluxes, computed by \citet[][priv. comm.]{Tacchella2023} using FORCEPHO (Johnson et al. in prep), are compared to the flux from each NIRCam image which falls within the MSA shutter. This is obtained by projecting the shutter corners on to the NIRCam images, accounting for distortion. 
The ratio between the two (the total flux to the apodized flux) can be used as a conversion factor. An additional wavelength-dependent correction of order $<$10\% is required because of additional flux losses due to additional diffraction by the MSA shutters, which leads to a broader PSF for NIRSpec than NIRCam at the same wavelength. This correction was approximated by the difference between the apodized image flux and the MSA through-shutter flux for a large sample of objects within the JWST Advanced Deep Extragalactic Survey (JADES) survey \citep{Bunker24, Boyett24}.  With the reversed point source correction, and applied morphology-sensitive slit loss and diffraction corrections, we then obtained the emission line fluxes used in this work's fitting (see Table \ref{tab:datatable}).  We will refer to \ticks{total equivalent} fluxes when referring to emission line fluxes that have been corrected according to the procedure described here.

In using the total image fluxes to correct the MSA through-shutter flux, we are assuming that the morphology of the regions emitting the emission lines follow that of the stellar continuum morphology at the wavelengths of the individual filters. In reality, the emission lines may have a very different morphology, since those originating from the NLR may have a much more compact morphology, and those from ionized gas around young stars may come from a clumpy distribution. 
 We discuss the impact of this assumption on our results in the Section \ref{subsec:aperture_corrections}.

\subsection{\NeIV\ Line Flux Extraction}
\label{subsec:fluxextraction}

For the extraction of the \NeIV\ flux, we used the 2D G235M/F170LP NIRSpec spectrum of SMACS S06355. The flux was summed across the $0.46^{\prime\prime}$ width of a micro-shutter (5 pixels), with the corresponding per-pixel noise added in quadrature in order to obtain measured flux error. After identifying the wavelength range occupied by the emission line, a straight line fit was made to the surrounding continuum, and subtracted from the flux. The rest-frame wavelength ranges considered to estimate the continuum level were 2406.00$\mathrm{\AA}$ - 2423.98$\mathrm{\AA}$ and 2429.71$\mathrm{\AA}$ - 2446.00$\mathrm{\AA}$ (blueward and redward of the peak, respectively). These ranges were taken to be featureless regions of the continuum.

To measure the line flux, we used the {\fontfamily{qcr}\selectfont emcee} python package to perform a
Bayesian fit assuming a Gaussian line shape and a Gaussian likelihood function, where we fitted the line flux (A), line centre (x$_0$) and line standard deviation ($\sigma$) over the prior ranges $0 < \mathrm{A}/ 10^{-17}\mathrm{erg}\,\mathrm{s}^{-1}\,\mathrm{cm}^{-2} < 0.45$, $2424  < \mathrm{x_{0}}/\mathrm{\AA} < 2428$ and $\mathrm{0.5 < \sigma/\AA < 2.5}$, respectively.

\begin{figure}
\includegraphics[trim={0 0.3cm 0 0},clip,width=0.5\textwidth]{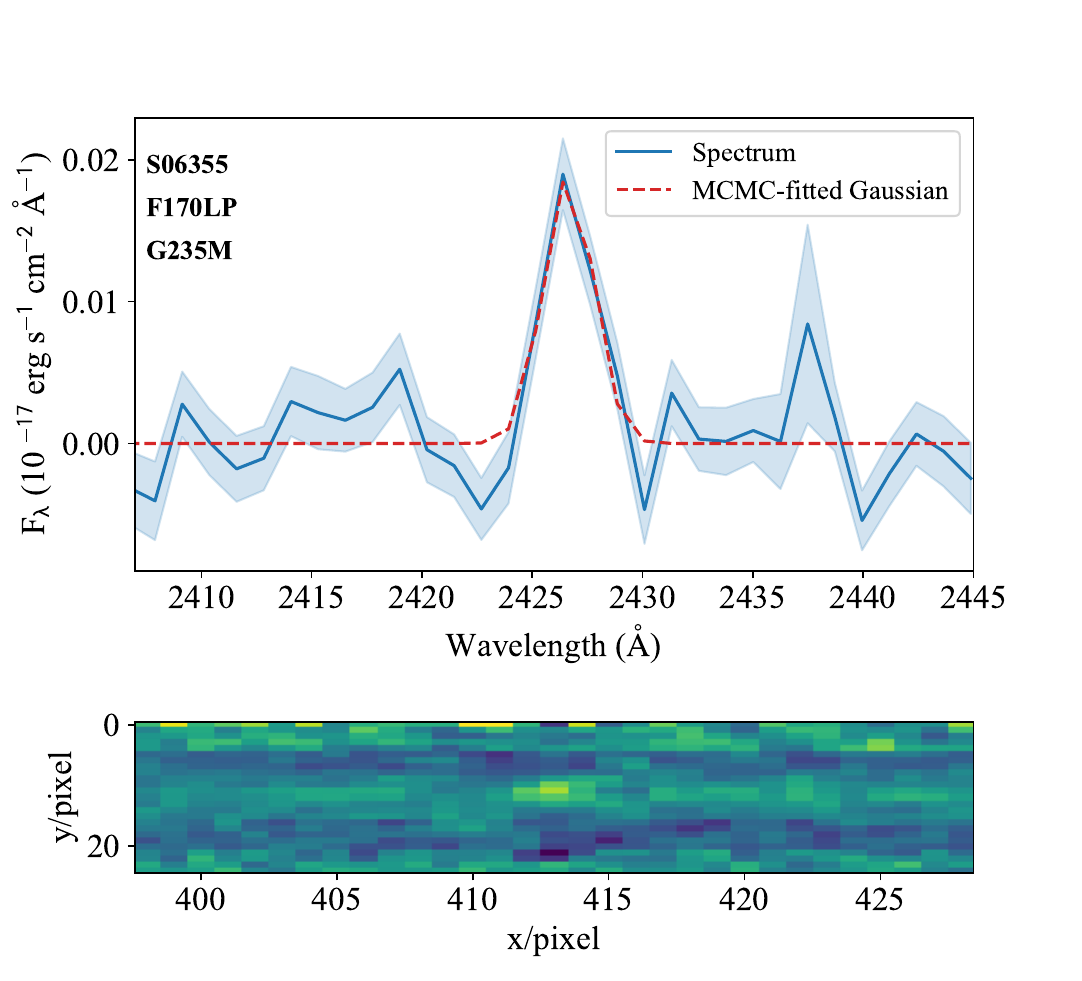}
\caption{The \NeIV\ emission from the F170LP/G235M spectrum of SMACS S06355. The lower panel shows the 2D spectrum, and the upper panel shows the 1D spectrum centred around the  \NeIV\ emission line. The measured spectrum is shown as a solid blue line, with light blue shading for the uncertainty and a dashed red line for the best-fitting model.}
\label{neivextraction}
\end{figure}

The MCMC process used 20 walkers and 8000 iterations. The observed, non-magnification corrected, \NeIV\ flux was measured to be $\mathrm{4.39 \pm 0.42 \times 10^{-19}}$ $\mathrm{erg \, s^{-1} cm^{-2}}$, and the observed spectrum and fitted Gaussian profile are shown in \newchange{Figure} \ref{neivextraction}. The significance of this emission line is approximately $10.5\sigma$; when incorporating the per-pixel noise across the wavelength dimension of the \NeIV\ emission itself, the significance is approximately $7.6\sigma$. We then pair the \NeIV\ flux we measure here with the following fluxes from table 1 of \cite{Curti2023a}: \OII, \Hdelta, \Hgamma, \OIIIft, \Hbeta, \OIIIfn, and \OIIIfi, and convert them to `total equivalent' line fluxes as described in Section \ref{subsec:data} and reported in Table~\ref{tab:datatable}. Prior to this, to ensure consistency between the different methodologies used for measurements of the line fluxes, we have also performed independent measurements of the emission lines using the same method as the \NeIV\ measurement, finding agreement within $\sim1.6\sigma$ on average.

\subsection{\normalfont\scshape{beagle-agn} \normalfont{Fitting}}
\label{subsec:beaglefitting}

As introduced in Section \ref{subsec:fluxextraction}, we \newchange{simultaneously} fit to both emission lines (\NeIV, \OII, \Hdelta, \Hgamma, \OIIIft, \Hbeta , \OIIIfn, and \OIIIfi) and filter fluxes (F090W, F150W, F200W, F277W F356W and F444W), using \beagleagn\ \citep{Vidal-Garcia2024}.  \newchange{The filter fluxes provide constraints on the e.g. the stellar mass, while the emission lines provide constraints on the gas (\HII\ and NLR) properties.} The emission lines in particular are included in the spectrum plotted in Fig. \ref{6355spectrum}, with the exception of \NeIV\ and \OIIIft; the former belongs to a different grating/filter spectrum (see Fig. \ref{neivextraction}) and the latter lies by the edge of the MSA chip gap and so is visually difficult to identify in Fig. \ref{6355spectrum} but was carefully extracted in \cite{Curti2023a}.

\beagle\ is an SED fitting code which models the stellar and nebular emission of galaxies. \beagleagn\ additionally includes the nebular emission from the narrow-line region (NLR) surrounding active galactic nuclei. The stellar models are an updated version of the \cite{bc03} stellar population synthesis models, where the updates are described in \citet[section 2.1 and appendix A]{Vidal-Garcia2017}. The NLR models are described in \cite{Feltre2016}, with updates regarding the addition of a smaller inner NLR radius and microturbulence in the clouds within the NLR described in \cite{Mignoli2019}.

\begin{table}
    \centering
    \caption{Prior limits, fixed values and other parameters used in this work's fit to SMACS S06355. Priors described with $\mathcal{N}{[\mu,\sigma^{2}]}$ notation denote a Gaussian profile with mean $\mu$ and standard deviation $\sigma$.}
    \begin{tabular}{c c}
    \toprule
       Parameter & Prior\\
       \midrule
        
        $\tauV$ & Exponential $\in [0,4]$ \\

        $\log(\Zhii/\Zsun)$ & Uniform $\in [-2.2, 0.24]$\\

        $\logUs$ & $\mathcal{N}[-2.5,0.75^{2}]$,  $\in [-4,-1]$\\

        $\log(\sfrInLog/\Msun \yr^{-1})$ & $\mathcal{N}[-0.0,2.0^{2}]$,  $\in [-4,4]$\\

        $\log(\ZAGN / \Zsun)$ & Uniform $\in [-2,0.3]$\\

        $\mathrm{\logUsAGN}$ & Uniform $\in [-4,-1]$\\

        $\mathrm{log(\Lacc/ erg s^{-1})}$ & Uniform $\in [43, 48]$\\

        $z$ & Fixed to 7.6643\\

        $\log(\MtotInLog/\Msun)$ & Uniform $\in [5,12]$\\
        
        $\log(\tausfr/\yr)$ & Uniform $\in [8,10.5]$\\

        $\log(t/\yr)$ & Uniform $\in [7,10]$ \\

        log($\mathrm{\Delta t_{SFR}/yr}$) & Fixed to 7\\
        
        $\mu$ & Fixed to 0.4\\
        
        $\xidAGN$ & Fixed to 0.3\\
        
        $\xid$ & Fixed to 0.3\\
        
        $\PLalpha$ & Fixed to -1.7\\

        $m_{\textrm{up}}$/$\mathrm{M_{\odot}}$ & Fixed to 100\\
        
    \bottomrule
    \end{tabular}
    
    \label{tab:fiducialparameters}
\end{table}

We account for dust attenuation with the two-component model of \cite{cf00}. This model assumes that radiation produced by young stars ($<10$ Myr) traverses the dusty birth clouds within which they are enshrouded before encountering the dust in the diffuse inter-stellar medium (ISM).  Older stars ($>10$ Myr), however, are only attenuated by the dust in the diffuse ISM. The age range of stars still enshrouded in their birth clouds approximates the lifetimes of the clouds in which the stars form, approximately $10^{7}$ yr (\citealt{Murray2010, Murray2011}). The attenuation experienced by the NLR region is accounted for in the modelling approach of \beagleagn\ in two ways. First, dust within the NLR itself is accounted for in the \cloudy\ models \citep[see][]{Feltre2016}, while light emitted from the NLR encounters further dust attenuation by the ISM \citep[for more details see][]{Vidal-Garcia2024}.  We test the dependence of our results on this simplified modelling of the NLR dust in Section~\ref{subsec:nlr_dust}.

We follow the prescription outlined in \cite{Gutkin2016} to compute the line and continuum emission of \HII\ regions ionized by stars younger than 10 Myr. We adopt a two-component star formation history.  The older component consists of a linear-exponential form described by $\sfrInLog=t\,\textrm{exp}(-t/\tausfr)$, where $t$ is the time since the first stars were formed, and \tausfr\ is the time-scale of star formation. This is followed by the most recent component, which is modelled as a constant star formation for a duration fixed to log($\mathrm{\Delta\t_{SFR} / yr}$) = 7, as indicated in Table \ref{tab:fiducialparameters}. The computation of stellar mass assembled during this two-component star formation history accounts for material returned to the ISM.  However, the integral of the star formation history, \Mtot, is the parameter that is sampled over, hence the prior given in Table \ref{tab:fiducialparameters} is over \Mtot\ rather than stellar mass. As with \beagle, \beagleagn\ can also adopt upper mass limits of either 100 $\mathrm{M_{\odot}}$ or 300 $\mathrm{M_{\odot}}$ for the \cite{Chabrier2003} IMF used. 

The spectrum of SMACS S06355 does not cover the full set of standard diagnostic BPT/VO87 (\citealt{Baldwin1981, Veilleux1987}) lines used to identify narrow-line region contribution at low redshift. Even if it did cover these emission lines, it is now known that BPT/VO87 diagrams generally misclassify galaxies at high redshift \citep[for example][fig. 8]{Maiolino2023b}, making the differentiation between different classifications more difficult with increasing redshift. \cite{Vidal-Garcia2024} found that fitting to line ratios when using the standard BPT/VO87 set of lines (plus \OII/\OIIIfi, \Hbeta/\Halpha\ and \OI/\OII) was required to obtain unbiased parameters. However, this was due to the lack of any high-ionization lines that could be attributed to the NLR alone. 
\newchange{The \NeIV\ emission line flux cannot be reproduced by the star-forming model grids (see Section \ref{subsec:alternativesources}) and is solely fit by the NLR models.  This provides valuable constraints on the NLR contribution to all the emission lines without resorting to line ratios.}
Therefore we fit to the fluxes of individual emission lines (in addition to photometric data), as opposed to ratios of emission lines as done in \cite{Vidal-Garcia2024}. In addition, since there are a large number of possible free parameters in \beagleagn\ but limited observations, we limit the number of AGN parameters we try to constrain.

The parameters primarily explored in our analysis are the $V$-band attenuation optical depth (\tauV), \HII\ region metallicity (\Zhii), \HII\ region ionization parameter ($\logUs^{\mathrm{HII}}$), star formation rate ($\psi$), NLR metallicity (\ZAGN), NLR ionization parameter (\logUsAGN) and accretion disc luminosity (\Lacc). We fix the hydrogen density of the NLR (\nHAGN $=$ 1000 cm$^{-3}$), the slope of incident ionising radiation to the NLR (\PLalpha $=$ -1.7)\footnote{\newchange{Power-law parameter \PLalpha\ is described further in Eqn. 5 ($\alpha$) of \cite{Feltre2016}.}}, the NLR covering fraction (10\%), the carbon-oxygen-ratio for each the \HII\ regions and NLR (C/O $=$ 0.44), and the fraction of the attenuation optical depth arising from the diffuse ISM (\mud $=$ 0.4). We also fix the \HII\ region hydrogen density to 100 cm$^{-3}$, and the dust-to-metal mass ratio of each component (\xid\ $=$ \xidAGN\ $=$ 0.3). The value choices of fixed parameters were guided by those within \cite{Vidal-Garcia2024}. Table \ref{tab:fiducialparameters} summarises these physical parameters and their corresponding priors used in the \beagleagn\ fitting process.

\begin{table*}
\begin{center}
\caption{\beagleagn\ fit for SMACS S06355, where $\log(\sfrInLog\newchangemath{/\Msun\,\mathrm{yr}^{-1})}$, $\newchangemath{\log(\xi_{ion})}$ and $\log(\MstarInLog/\Msun)$ from this work have been corrected for magnification ($\mathrm{\mu = 1.23 \pm 0.01}$ as used in \citealt{Curti2023a}, originating from the lens model of \citealt{Mahler2023}), and are quoted with uncertainties corresponding to the 68\% credible interval. We present our results alongside those from previous studies, namely those of \protect\cite{Tacchella2023} and \protect\cite{Curti2023a}. Cells with a \ticks{$-$} denote values which are either not applicable to the study, or otherwise not attributed.}
\begin{tabular}{ C{0.3\columnwidth}C{0.3\columnwidth}C{0.3\columnwidth}C{0.3\columnwidth}} 
\toprule

 \multicolumn{1}{c}{Parameter} &
 \multicolumn{1}{c}{This Work} & \multicolumn{1}{c}{\cite{Tacchella2023}} & \multicolumn{1}{c}{\cite{Curti2023a}}\\
 
\midrule

        \tauV &  $\newchangemath{0.84^{+0.34}_{-0.28}}$ &  $0.43^{+0.26}_{-0.15}$ & $0.50^{+0.26}_{-0.15}$\\

        $\log(\Zhii/\Zsun)$ & $\newchangemath{-0.89^{+0.18}_{-0.19}}$ & $-0.6^{+0.1}_{-0.1}$ & $-$\\

        \logUs &  $\newchangemath{-2.06^{+0.13}_{-0.14}}$ & $-$ & $-$ \\

        $\log(\sfrInLog/\Msun\,\textrm{yr}^{-1})$ &  $1.70^{+0.10}_{-0.09}$ & $1.49^{+0.14}_{-0.09}$ & $1.47^{+0.04}_{-0.04}$\\

        $\log(\xi_{ion} / \mathrm{erg Hz^{-1}})$ &  $26.11^{+0.11}_{-0.10}$ & $-$ & $-$ \\

        $\log(\MstarInLog/\Msun)$ &  $9.12^{+0.16}_{-0.16}$ & $8.60^{+0.20}_{-0.20}$ & $8.72^{+0.04}_{-0.04}$ \\

        $\log(\ZAGN/\Zsun)$ & $\newchangemath{0.12^{+0.14}_{-0.14}}$ & $-$ & $-$ \\

        \logUsAGN &  $\newchangemath{-1.93^{+0.25}_{-0.23}}$ & $-$ & $-$\\

        $\log (\Lacc/\mathrm{ergs^{-1}})$ & $45.19^{+0.12}_{-0.11}$ & $-$ & $-$ \\

        $\mathrm{[OIII]\lambda4363^{HII}}$/\%  & $\newchangemath{57.70^{+9.52}_{-9.60}}$ & 100 & 100\\    

        $\mathrm{H\beta\lambda4861^{HII}}$/\%  & $\newchangemath{75.60^{+8.06}_{-6.93}}$ & 100 & 100\\
        
        $\mathrm{[OIII]\lambda5007^{HII}}$/\% & $\newchangemath{49.08^{+12.95}_{-11.71}}$ & 100 & 100\\

        $^{1}$12+log(O/H)$^{\mathrm{HII}}$ &  $\newchangemath{7.83^{+0.18}_{-0.19}}$ & $-$ & $8.24^{+0.07}_{-0.07}$\\

        $^{1}$12+log(O/H)$^{\mathrm{NLR}}$ & $\newchangemath{8.86^{+0.14}_{-0.16}}$ & $-$ & $-$ \\

        $\mathrm{\chi^{2}_{min}}$ & 27.85 & $-$ & $-$ \\        

\bottomrule
\end{tabular}\label{tab:resultstable}
\end{center}
\vspace{0.1cm}
{\footnotesize $^{1}$\newchange{These values represent tracers of the metallicity log($\Z/\Zsun$) which includes all metals; in this format it is the gas-phase abundance which depends on a combination of log($\Z/\Zsun$) and dust-to-metal mass ratio \xid}.}
\end{table*}

\section{Results}\label{section:results}

Our \beagleagn\ parameter estimates are detailed in Table~\ref{tab:resultstable}. Within this table are additional columns including results from previous studies of SMACS S06355 (\citealt{Tacchella2023} and \citealt{Curti2023a}). Comparisons between these previous studies and this work will be discussed in Section \ref{subsec:compstoothers}. Reported parameter estimates from this work have been magnification corrected according to the lens model of SMACS J0723.3-7323 from \cite{Mahler2023}.

\begin{figure*}
\includegraphics[width=1\textwidth]{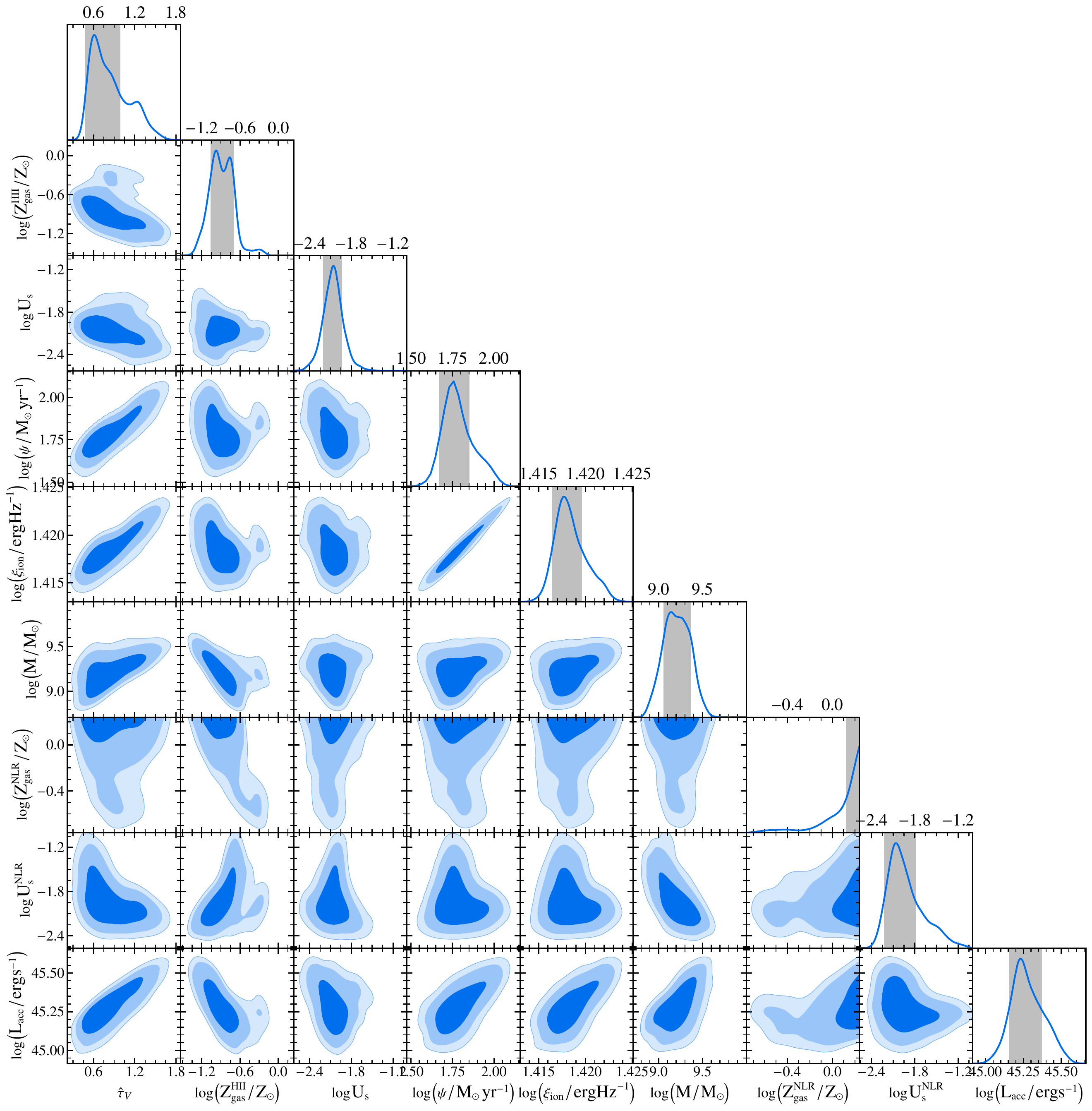}
\caption{2D posterior probability distributions generated from this work's \beagleagn\ fit to SMACS S06355. Central dark blue contours refer to the 1$\sigma$ confidence level, whereas the second and third concentric contours reflect the 2$\sigma$ and 3$\sigma$ levels respectively. The uppermost panels for each column display the 1D posterior probability distributions for their corresponding parameters, where vertical gray shading denotes the 1$\sigma$ confidence interval. Parameters shown are $V$-band attenuation optical depth $\hat{\tau}_{v}$, \HII\ region metallicity log(\Zhii / \Zsun), \HII\ region ionization parameter \logUs, star formation rate $\log(\sfrInLog/\Msun\yr^{-1})$, ionising emissivity log($\xi_{\mathrm{ion}} / \mathrm{erg\ Hz^{-1}}$), stellar mass $\log(\MstarInLog/\Msun)$, AGN metallicity log(\ZAGN/Z$_{\odot}$), AGN ionization parameter \logUsAGN, and AGN accretion luminosity log($\mathrm{\Lacc / ergs^{-1}}$). Quantities with \ticks{log} refer to base 10 here and throughout.}
\label{triangleplot}
\end{figure*}

\begin{figure*}
\includegraphics[width=0.85\textwidth]{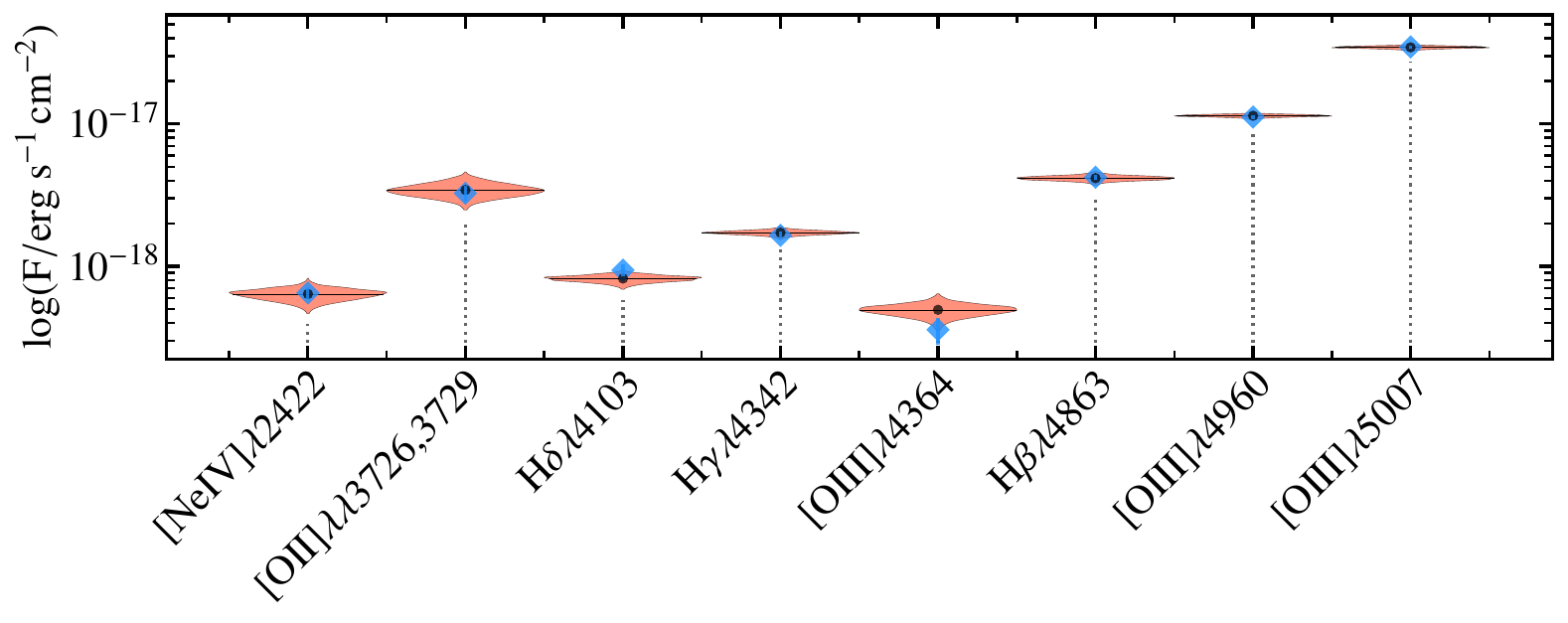}
\caption{A \pypbeagle\ violin plot showing the comparisons between the measured (blue filled diamonds) and modelled (black filled circles) fluxes for corresponding emission lines associated with this work's \beagleagn\ fit to SMACS S06355. Here, the tomato shading represents the probability distributions of the modelled fluxes. The solid blue bars show the measured flux errors, and are occasionally smaller than the blue diamond markers; errors are visualised more clearly in Fig. \ref{chiplot}.}
\label{marginalplot}
\end{figure*}

\begin{figure}
\includegraphics[trim={1.45cm 1.45cm 0.25cm 0.25} , width=0.5\textwidth]{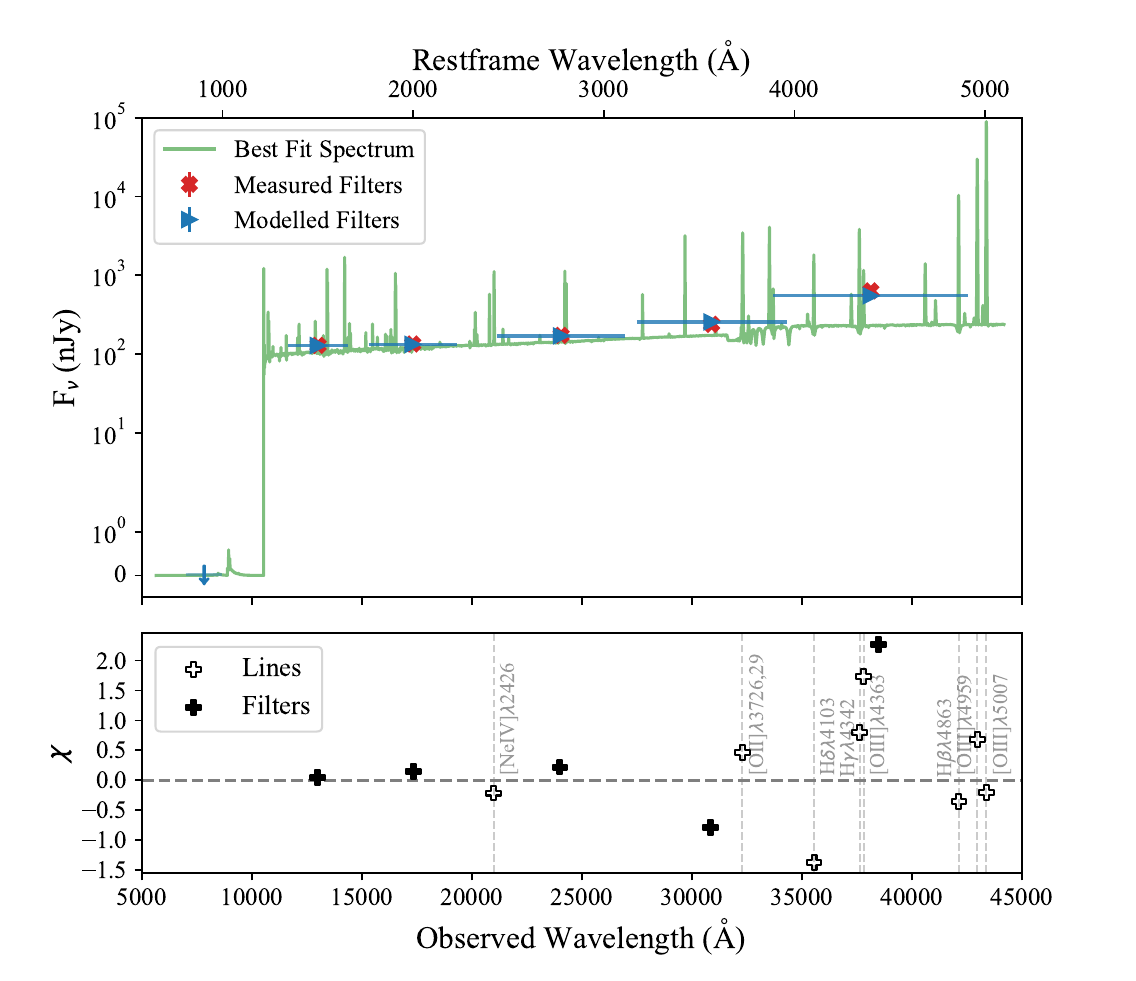}
\caption{The \beagleagn\ fit to SMACS S06355. The upper panel shows the best-fitting spectrum (light green solid line), over-plotted with modelled (blue filled rightward pointing triangles) and measured (red filled crosses) filter fluxes. The F090W modelled filter flux is presented as an upper limit (blue downward pointing arrow). The bandwidths of each filter throughput curve are visualised as horizontal error bars. The lower panel includes different $\chi$ values, namely those corresponding to the filter fluxes (solid black plus markers) and also the emission line fluxes (hollow black plus markers).}
\label{chiplot}
\end{figure}

The 2D posterior distributions for the parameters are plotted in \newchange{Figure} \ref{triangleplot}; these present derived parameters which are mostly reasonably constrained considering the redshift of this source and the number of data measurements used in the fitting. Caveats to this typically include AGN related parameters (namely NLR metallicity $\mathrm{log(\ZAGN / \Zsun)}$, see discussion in Section \ref{subsec:HIINLRmets}), which is to be expected in this specific case of having only one very high-ionization potential line ($\NeIV$) to constrain NLR parameters without contamination from star formation. The plot in \newchange{Figure} \ref{marginalplot} shows a comparison between the measured and modelled emission line fluxes. \newchange{Figure} \ref{chiplot} visualises the best-fitting model spectrum of SMACS S06355 and the statistical quality of the fit. Together, Figs. \ref{marginalplot} and \ref{chiplot}
demonstrate the consistency between modelled emission line fluxes and their corresponding measured values ($\mathrm{\overline{\chi_{lines}^{2}} = 0.80}$). In Fig. \ref{chiplot}, we see the modelled filter flux values also reproduce the measured values reasonably well with $\mathrm{\overline{\chi_{filters}^{2}} = 1.17}$.

With \beagleagn, we can additionally explore the fractional contribution of the \HII\ and narrow-line regions to various emission lines; examples of such contributions are presented in the \ticks{$\mathrm{H\beta\lambda4861^{HII}}$} and \ticks{$\mathrm{[OIII]\lambda5007^{HII}}$} rows of Table \ref{tab:resultstable}.

Massive stars and top-heavy IMFs have been proposed as likely candidates for explaining high N/O ratios and other observables seen at high redshift (e.g. \citealt{Cameron2024, Vink23}). In the case of this work, it is desirable to explore the impact of changing the IMF upper \newchange{mass} limit on our results; whilst we do not fit with a top-heavy IMF specifically, \beagleagn\ allows us to increase the IMF upper mass limit to 300\Msun, which we explore without changing any of the other fit parameters. In this fit, the SFR decreased by approximately 0.17 dex as we may expect; this is due to more massive stars having higher luminosities than lower mass stars, and so requiring less star formation to produce the same fluxes for a given time period. No parameter changed by more than 1$\sigma$ relative to the fit presented in Tab. \ref{tab:resultstable} apart from the SFR and nebular ionization parameter, which have decreased by 1.70$\sigma$ and 1.71$\sigma$ respectively. Therefore, overall our 100\Msun\ and 300\Msun\ results are not statistically different. When we perform a fit identical to our fiducial, but without the AGN component (so SF only), we derive a higher SFR of log($\sfrInLog/\Msun\ \mathrm{yr^{-1}}$) $= 1.83^{+0.03}_{-0.03}$ in comparison to our fiducial value, but the \NeIV\ flux is not reproduced by the model. 

Considering our results given in Table \ref{tab:resultstable}, SMACS S06355 hosts a moderately luminous Type-II AGN (log(\Lacc/$\mathrm{ergs^{-1}}$) = $45.19^{+0.12}_{-0.11}$). In this model, the \HII\ region metallicity appears significantly sub-solar (log(\Zhii/\Zsun) = $\newchangemath{-0.89^{+0.18}_{-0.19}}$), while the NLR metallicity posterior pushes towards higher values though the full posterior is relatively unconstrained in this case:  log(\ZAGN/\Zsun) = $\newchangemath{0.12^{+0.14}_{-0.14}}$ (discussion expanded in Section \ref{subsec:HIINLRmets}). Both metallicity estimates show large uncertainties, primarily because of the single very high-ionization line (\NeIV) used in the analysis and sensitive to the AGN conditions, for the case of NLR metallicity. We note that the derived NLR metallicity, in particular, is likely to be dependent on the gas-phase Neon abundance in the models, and we discuss this further in Section~\ref{subsec:NeIII}.

\section{Discussion}\label{section:discussion}

\subsection{Comparison to other studies}
\label{subsec:compstoothers}

Within Table \ref{tab:resultstable} are results of this work which we can compare to some of the previous studies of SMACS S06355, namely those of \cite{Tacchella2023} and \cite{Curti2023a}. 
This is the first work that simultaneously accounts for \HII\ and the narrow-line region contributions to the emission lines (which can introduce degeneracies). 
Additionally, this work includes the \NeIV\ emission line, not adopted in the other works, and adopts data including `total equivalent' line fluxes as outlined in Section \ref{subsec:data}. Therefore, whilst we perform comparisons to previous studies as outlined, we do so whilst recognising this work's approach is innately different.

The optical depth in the $V$-band from this work, \tauV\ $=$ $\newchangemath{0.84^{+0.34}_{-0.28}}$, despite having a higher mean is not statistically different to the values derived from previous works, which found $0.50^{+0.03}_{-0.03}$ and $0.43^{+0.26}_{-0.15}$ for \cite{Curti2023a} and \cite{Tacchella2023}, respectively. We discuss the treatment of and dependence of dust further in Section \ref{subsec:nlr_dust}.

Results from this work indicate a galaxy with an intense SFR $\mathrm{\sfr = 51.12^{+12.98}_{-9.38}}$ $\mathrm{M_{\odot}}$ $\mathrm{yr^{-1}}$.  We also note the SFR here is higher than those reported in previous studies of the same source with $29.51^{+2.85}_{-2.60}$ and $31^{+12}_{-6}$ $\mathrm{M_{\odot}}$ $\mathrm{yr^{-1}}$ for \cite{Curti2023a} and \cite{Tacchella2023}, respectively. When comparing SFRs of SF and SF+NLR fits, if both use the same input data we would intuitively expect the SF+NLR fit to derive a lower SFR than the SF only fit. Indeed, the SF only version of this work's fit has a higher SFR than the SF+NLR version ($\sim67$ compared to $\sim51$ \Msun yr$^{-1}$). However, here we largely attribute the opposite behaviour observed to the higher derived dust attenuation optical depth \tauV, which has likely driven the SFR higher in our fitting. Differences in stellar models and star formation histories employed may also introduce differences in these inferred SFRs. For example, the SFR reported in \cite{Tacchella2023} is that averaged across their initial two time bins (equivalent to 10 Myr), as part of a 6 bin flexible SFH prescription using a bursty prior. In comparison, our fitting adopts a delayed SFH with an  added 10 Myr variable star-forming period of constant SFR.  

We can also compare this work's derived \HII\ metallicity to those found in \cite{Curti2023a} and \cite{Tacchella2023}. Our \HII\ region metallicity, log(\Zhii/\Zsun) = $\newchangemath{-0.89^{+0.18}_{-0.19}}$,  is approximately $2\sigma$ away from that obtained in \citet[log(\Zhii/\Zsun) = $-0.60^{+0.10}_{-0.10}$]{Tacchella2023}. We can also compare our fiducial \HII\ region gas-phase metallicity, $\mathrm{12 + log(O/H)^{HII}} = \newchangemath{7.83^{+0.18}_{-0.19}}$, to that of \citet[$\mathrm{12 + log(O/H)^{HII}} = 8.24^{+0.07}_{-0.07}$]{Curti2023a} in a similar manner. Here, our fiducial value is approximately $\newchangemath{2.3}\sigma$ away, where the value from \cite{Curti2023a} is being derived from the direct T$\mathrm{_{e}}$ method and assuming that \textit{all} line emission is powered by stellar emission. In contrast, ours is derived from comparison to photoionization models while dis-entangling the contribution from the \HII\ and NLR regions, and so deriving a comparatively lower gas phase metallicity here is what we may expect with our dual component approach. The derived \HII\ and NLR metallicities from this work are not highly constrained, and their posterior distributions within Fig. \ref{triangleplot} present generally non-gaussian forms. Therefore, when comparing our derivations to previous works, quoting posterior means and uncertainties potentially omits additional information. A way we mitigate this is exemplified in \newchange{Figure} \ref{metallicity_plot}, where we plot the posterior distributions themselves within an appropriate shared parameter space, as Section \ref{subsec:HIINLRmets} expands.

\subsection{\HII\ and NLR metallicities}
\label{subsec:HIINLRmets}

\begin{figure*}
\centering
\includegraphics[width=1.0\textwidth]{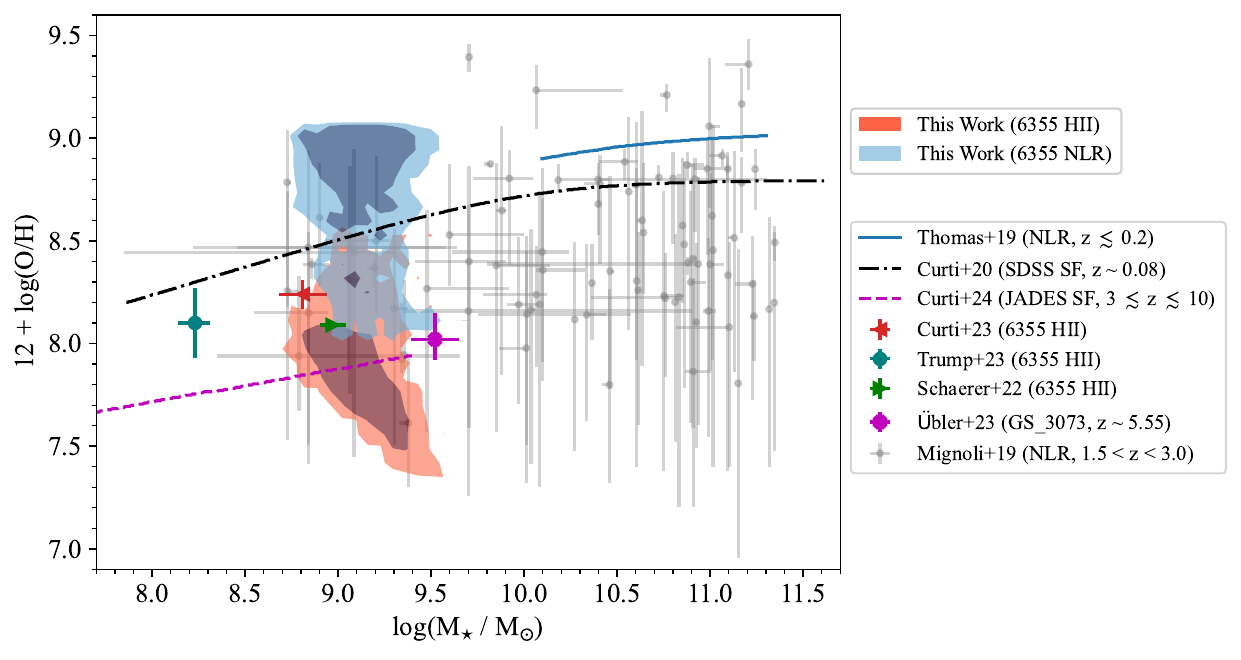}
\caption{The Mass-Metallicity Relation (MZR), compiling various works, where all studies of SMACS S06355 ($\mathrm{z \sim 7.66}$) are indicated with a \ticks{6355} label. This work presents metallicities from all live points of our \beagleagn\ fit of both the NLR (blue filled contours) and \HII\ region (red filled contours). The inner and outer contours contain 68\% and 95\% of the associated probability, respectively\newchange{, and the stellar mass values have been magnification corrected assuming }$\newchangemath{\mu = 1.23}$. Additional estimates for the \HII\ region metallicity of SMACS S06355 include those of \protect\citet[red left-pointing triangle]{Curti2023a}, \protect\citet[large blue circle]{Trump2023} and \protect\citet[green right-pointing triangle]{Schaerer2022}. We additionally make comparisons to the 1.5 < z < 3.0 NLR AGN from \protect\citet[small gray circles]{Mignoli2019}; the $\mathrm{z \lesssim 0.2}$ obscured AGN based MZR from \protect\citet[solid blue line]{Thomas2019}, the $\mathrm{z \sim 0.8}$ SDSS star-forming galaxies from \protect\citet[dot-dashed black line]{Curti2020} and the $\mathrm{3 \lesssim z \lesssim 10}$ low stellar mass  JADES star forming sources from \protect\citet[dashed magenta line]{Curti2023b}. For reference, this MZR compilation also includes the $\mathrm{z \sim 5.55}$ Type-I AGN from \protect\citet[magenta cross]{Ubler2023}.}
\label{metallicity_plot}
\end{figure*}

As seen in Section \ref{subsec:compstoothers}, the \HII\ region metallicity that we derive is lower than the one in previous studies, likely a result of the different methods used. The posteriors from our derived \HII\ and NLR gas-phase metallicities for this object are visualised on the mass-metallicity plane in Fig. \ref{metallicity_plot}. This figure also includes results from studies of obscured AGN and star-forming galaxies, which allows us to consider the position of SMACS S06355 in relation to these.

Our \HII\ region gas-phase metallicity falls below the SDSS mass-metallicity relation \citep{Curti2020} by approximately $\newchangemath{0.67}$ dex (taking the means of the $\mathrm{12 + log(O/H)^{HII}}$ and $\mathrm{log(M/M_{\odot})}$ posterior PDFs), as we may expect from metallicity trends at higher redshifts \citep{MadauDickinson2014}. Here, the derived \HII\ MZR posterior, as part of the \beagleagn\ fit, is consistent with the MZR from the low mass star-forming JADES sources across $3 < z < 10$ \citep{Curti2023b}. The previous estimates of SMACS S06355's (HII region) position on the MZR, apart from that of \cite{Trump2023}, are in approximately 1$\sigma$ proximity to this work's 95\% \HII\ probability region. In the case of \cite{Trump2023}, the stellar mass was adopted from \cite{Carnall2023}, therein derived using \textsc{bagpipes} \citep{Carnall2018}.  

The 68 and 95 per cent posterior probability regions  from this work's NLR characterisation are also shown in Fig. \ref{metallicity_plot}, and can be compared to other NLR mass-metallicity estimates across different redshifts. In Fig. \ref{metallicity_plot}, the MZR attributable to the $z \lesssim 0.2$ Type-II sources from SDSS (Sloan Digital Sky Survey) as presented in \cite{Thomas2019} is shown as a solid blue line.  \cite{Thomas2019} used \textsc{NebularBayes}, which can also simultaneously disentangle the \HII\ and NLR contribution to emission lines, though they employ different nebular emission models. In their study, the derived NLR MZR relation is based on a fitting to a mixed functional form of the MZR from \cite{Moustakas11}, where two of three free parameters were fixed to values attributed to a fit to the star-forming galaxies in the same paper \citep{Thomas2019}. This was appropriate as the SF galaxies populated a large enough stellar mass range to constrain all three parameters, in comparison to the AGN objects. In recognition of this, we refrain from extrapolating the \cite{Thomas2019} relation to the stellar mass of SMACS S06355. Generally, we observe that the 68\% probability region of our NLR mass-metallicity covers the range of metallicities traversed by the \cite{Thomas2019} fitted MZR, though the fit also allows for significantly lower values. 

Fig. \ref{metallicity_plot} also includes the mass and metallicity estimates for $\mathrm{1.5 < z < 3.0}$ CIV$\lambda$1549-selected Type-II AGN presented in \cite{Mignoli2019}.  The results from this study generally lie below our derived NLR metallicity. This study applied a two-phase (galaxy + AGN) SED fitting technique, further described in \cite{Bongiorno2012}, to optical and near-infrared photometry to derive the stellar masses. These stellar masses were paired with NLR metallicities to explore objects on the MZR plane. The NLR metallicities from \cite{Mignoli2019} were obtained by fitting emission line ratios using the \cite{Feltre2016} prescription, with added adjustable parameters of gas cloud internal microturbulence velocity, and NLR inner radius. They assumed the full line flux was attributable to the NLR, and fit only with rest-UV emission lines covering medium-high ionization potentials \citep[see][tab. 1]{Mignoli2019}. Indeed, the selection criteria of the sample itself may impact its coverage of the mass-metallicity plane.  We do not have coverage of \CIV\ to know whether this object would have met their sample selection criteria (though see discussion in sections ~\ref{subsec:NeIII} and ~\ref{subsec:aperture_corrections} for possible reasons for high NLR estimations from our modelling). 
 
In addition to comparisons with other studies of obscured AGN on the MZR, we can compare to studies of Type-I AGN. For example, our NLR metallicity (\ZAGN/\Zsun $\mathrm{= \newchangemath{0.12^{+0.14}_{-0.14}}}$ , 12 + log(O/H)$\mathrm{^{NLR}}$ = $\newchangemath{8.86^{+0.14}_{-0.16}}$) 68\% posterior region occupies a higher metallicity space than the NLR metallicity of the $z \sim 5.55$ Type-I AGN presented in \citet[$\mathrm{8.00^{+0.12}_{-0.09}}$]{Ubler2023}. It also appears consistent with the typical metallicity estimates of z > 4 BLR AGN, namely those of \citet[$\mathrm{Z\ \sim\ 0.2\ \Zsun}$]{Maiolino2023b}. These metallicity inferences were based on comparisons between the space occupied by the BLR AGN on a BPT/VO87 diagram \citep{Baldwin1981,Veilleux1987}, and the space occupied by the AGN photoionization models of \cite{NakajimaMaiolino2022}.

\subsection{Alternative sources of ionising radiation} \label{subsec:alternativesources}

The grounds for SMACS S06355 being an obscured AGN largely lie in the \NeIV\ emission line in its F170LP/G235M spectrum \citep{Brinchmann2023}. The ionization potential of the \newchange{Ne$^{2+}$ ion} ($\sim$ 63 eV) is beyond what standard stellar populations can achieve. For example, the SF photoionization models of \cite{Gutkin2016} produce \NeIV\ equivalent widths (EW) $\lesssim 0.2 \AA$, approximately 60 times lower than our measured value for SMACS S06355, indicating that the \NeIV\ line originates from a source other than star formation. In the context of this work's SED fitting, we investigated this by performing a fit identical to that in Section \ref{section:results}, but without an AGN component. In this fit, the modelled \NeIV\ flux failed to reproduce the measured value ($\chi_{\mathrm{[NeIV],100}} = 10.39$), indicating that star formation alone is unable to account for the observed \NeIV.

Whilst looking for high-ionization emission lines as signatures of AGN activity is advantageous when local diagnostics become unreliable, this approach cannot currently  eliminate the possibility of such radiation originating from other highly ionising sources conclusively. Literature has shown that models including massive stars and  top-heavy IMFs have been able to reproduce observed high ionising potential lines in certain cases (e.g. \citealt{Cameron2024, Lecroq24, Topping2024}). Given \textit{z} > 6 stellar populations are typically thought to host stars with very high masses, it is sensible to consider a top-heavy IMF as an alternative source of the high ionising potential lines in the spectrum of SMACS S06355, namely \NeIV. We explore this by performing a fit identical to that presented in Section \ref{section:results}, but with no AGN component and an IMF upper mass limit of 300\Msun\ as opposed to 100\Msun. This fit still failed to reproduce the measured \NeIV\ line ($\chi_{\mathrm{[NeIV],300}} = 10.39$), indicating a lack of reproducibility even when increasing the IMF upper mass limit. A caveat to this at present is that a higher IMF upper mass limit is not necessarily equivalent to a top-heavy IMF slope, which maintains a slight uncertainty in our conclusion. Though we cannot explore this with more extreme populations of massive stars with our current model sets, the binary model described within \cite{Lecroq24} includes merged binary stellar masses between 2 and 600 $\Msun$. Their figure 2 shows that the ionising flux still falls short of the ionization potential for \NeIV.

Wolf Rayet (WR) stars also appear to be possible origins of high ionising potential lines (and other unique observables) in the early Universe. For instance, the strong N/O abundance found in GN-z11 \citep{Bunker23}, linking to the observed \NIV\ line (ionising potential $\sim$ 47 eV), was shown to be explainable by the chemical evolution models of \cite{Kobayashi2024}, in concert with SFHs with both quiescent and bursty phases, with stellar populations including WR stars. Indeed, if such stars are not uncommon at these early epochs, their supernovae may inject abundances of Neon in to the ISM. Within this work, our fitting approach makes use of models which include WR prescriptions \citep{Vidal-Garcia2017}, and so with our previous testing of \NeIV\ reproducability using SF models only, we similarily conclude WR stars are likely not the dominant ionising source of \NeIV.

Another alternative source of the high ionising potential line \NeIV\ includes X-ray binary stars (XRBs). XRBs have already been proposed as plausible origins of unusual emission line ratios. For example, \cite{Katz2023} show that including high mass XRBs in their modelling reasonably reproduced the high \OIIIft/\OIIIfi\ flux ratio of SMACS S04590, to a better extent than their AGN consideration. In comparison to \cite{Katz2023}, the emission line of interest in SMACS S06355 (\NeIV) \newchange{indicates the presence of high energy photons} ($\gtrsim$63 eV). XRBs still remain a valid alternative to an obscured AGN since additional astrophysical ingredients of XRBs  are typically able to account for high-ionization emission lines (e.g. \citealt{Schaerer2019, Lecroq24}), again resolving a specific need for AGN contribution. For instance, \cite{Garofali2024} found that a joint model combining a simple stellar population (SSP) with a simple X-ray population incurred intensity increases of emission lines such as $\mathrm{[NeV]\lambda3426}$ (ionising potential $\sim$ 97.19 eV), in comparison to SSP only models. If XRBs can enhance emission lines with ionising potentials $\mathrm{> 90}$ eV, then it is sensible to consider the same being possible for emission lines $\mathrm{< 90}$ eV also, for instance \NeIV. However, during the discussion of \citet[see their section 5]{Lecroq24}, it was highlighted that the claims of XRBs contributing significantly to high ionising potential and UV emission lines are typically based on models which predict improbably high X-ray luminosity - SFR ratios. From this discussion, though the XRB solution to an emission line like \NeIV\ appears tentative, we recognize it to be none the less more likely than the previously discussed alternative dominant sources at present  
 
We recognize that there may remain alternative origins for the observed \NeIV\ emission line. Though exploring this quantitatively is outside the scope of this current paper, with our discussion we consider the strongest alternative for the observed \NeIV\ in S06355 to be X-ray binary stars, if not a Type-II AGN (for which we regard most likely, and present this work's characterisation of SMACS S06355).

\subsection{Bolometric luminosity}
\label{subsec:bolometricluminosity}

With our results, it is also possible to explore the bolometric luminosity of the AGN. To compare to other bolometric luminosities obtained in the literature, we use an empirical calibration between \OIIIfi\ and \Hbeta\ luminosities, and the AGN bolometric luminosity as presented in \cite{Netzer2009}:

\vspace{0.25cm}
\begin{equation}
\begin{aligned}
    \mathrm{L_{[bol]}} = \mathrm{logL(H\beta)} + 3.48 + \mathrm{max}\left[0.0,0.31\left(\mathrm{log}\frac{\mathrm{[OIII]}}{\Hbeta} - 0.6\right) \right] \\
\label{netzer_lum}    
\end{aligned}
\end{equation} 
\vspace{0.1cm}

\noindent Within \beagleagn, it is possible to distinguish between \HII\ and NLR fractional contributions to emission lines. Therefore, when using Equation \ref{netzer_lum}, we can use the fractional contributions to \Hbeta\ and \OIIIfi\ from the NLR component, and achieve an estimate of $\mathrm{L_{bol}}$ which is more decoupled from star formation contributions. Using our fit results whilst correcting for magnification and dust attenuation, our current estimate becomes $\mathrm{log(L_{bol} / \mathrm{ergs^{-1}}) = 45.67 \pm 0.33}$, which is statistically similar to our derived log($\Lacc / \mathrm{erg s^{-1}}$) = $45.19^{+0.12}_{-0.11}$. Our $\mathrm{L_{bol}}$ estimate is approximately 0.53 dex smaller than the bolometric luminosity estimate found for the z $\approx$ 5.55 Type-I AGN described in \citet[$\mathrm{log(L_{bol, Netzer} / \mathrm{erg s^{-1}}) \approx 46.2}$]{Ubler2023}. 
Considering uncertainties, it is at the high end of the range probed by the $4 < z < 11$ Type-I AGN presented in \citet[$\mathrm{log(L_{bol, Netzer} / \mathrm{erg s^{-1}})}\sim 43.8-45.6$]{Maiolino2023b}. Our estimate of $\mathrm{L_{bol}}$ is therefore comparable to typical bolometric luminosity estimates of AGN at high redshifts.

\begin{figure}
\includegraphics[width=0.45\textwidth]{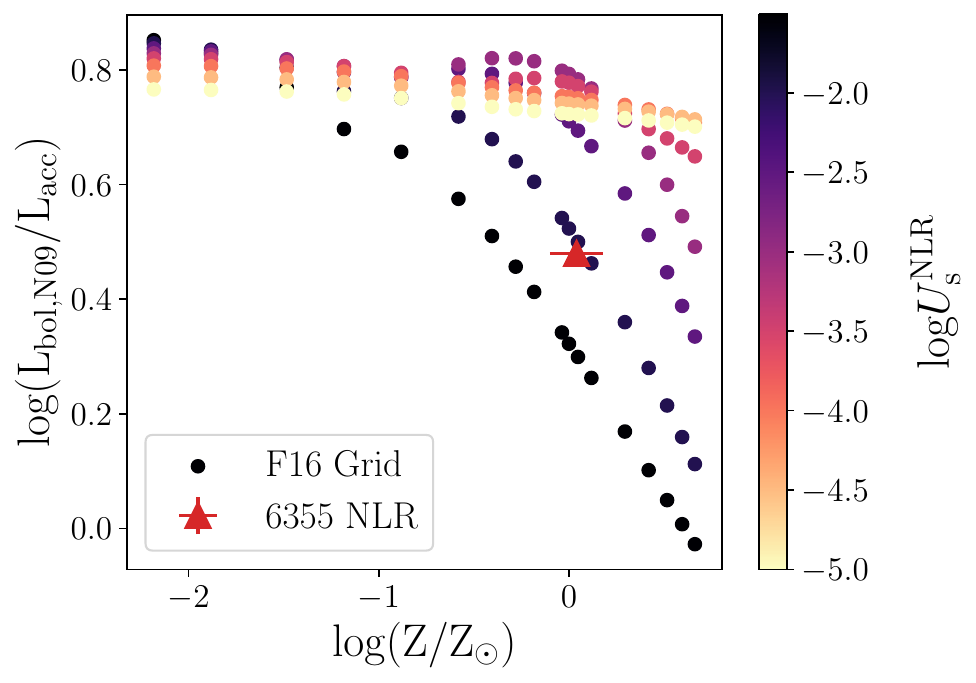}
\caption{Logarithm of the ratio between the \protect\cite{Netzer2009} bolometric luminosity calibration (see their equation 1) compared to the accretion luminosity for \protect\cite{Feltre2016} models plotted against metallicity for a wide range of ionization parameters.  For this figure we only plot models with \PLalpha\ fixed to $-1.7$, \nHAGN\ of 10$^3$ cm$^{-3}$ and \xidAGN\ = 0.3. The \protect\cite{Feltre2016} grid points are coloured according to their \logUsAGN\ values, and this work's SMACS S06355 point is overplotted as a red triangle ($\logUsAGN=-2.02^{+0.25}_{-0.23}$).}
\label{fig:NetzerComparison_1}
\end{figure}

\begin{figure}
\includegraphics[width=0.50\textwidth]{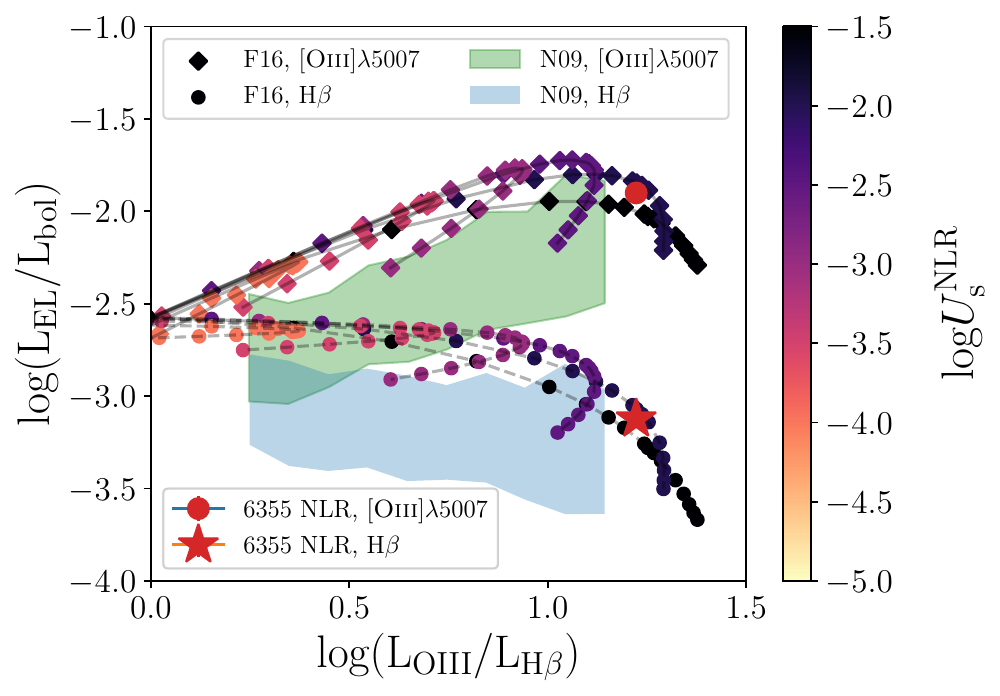}
\caption{Logarithm of the ratio between a given emission line luminosity (\OIIIfi\ or \Hbeta) and the bolometric luminosity plotted against the log-ratio of \OIIIfi\ and \Hbeta\ luminosities, a proxy for ionization parameter. The values for the \protect\cite{Feltre2016} models are shown as symbols connected by lines. Different values for ionization parameter are indicated by the colour bar, while the range of metallicity, values between $-2.2<\log(\ZAGN/\Zsun)<0.4$, increase from left to right until the relationships turn over and back on themselves at the highest metallicities. The green and blue shaded regions cover the areas defined by the mean and standard deviations of samples of measurements taken from Type-I AGN in the \protect\cite{Netzer2007} sample as presented in fig. 3 of \protect\cite{Netzer2009}, for \OIIIfi\ and \Hbeta\ respectively. Results for this work, specifically the \OIIIfi\ and \Hbeta\ related luminosity comparisons for SMACS S06355, are indicated by large red circle and star symbols, respectively.}
\label{fig:NetzerComparison_2}
\end{figure}

In principle, the parameter \Lacc\ is equivalent to the bolometric luminosity of the AGN and we could compare our derived \Lacc\ values to bolometric luminosity measurements of other sources (in all cases).  Despite the identified consistency between our derived \Lacc\ and $\mathrm{L_{bol}}$ in this case, we still also recognize that \Lacc\ is derived from the integral of a simple functional form, only a small fraction of which is constrained from the emission line data to hand \citep[see fig. 1 of][]{Vidal-Garcia2024} and so there can be many uncertainties in this comparison in other instances. To investigate these further, we compare the bolometric luminosity we would derive from the models using the \cite{Netzer2009} conversion (Fig.~\ref{fig:NetzerComparison_1}) for the model grid with fixed $\Lacc=10^{45}\,\textrm{erg}\,\textrm{s}^{-1}$. We see that there is a fairly stable comparison at low metallicities ($\log(\Z/\Zsun)<-1$), albeit with a 0.8 dex offset to higher bolometric luminosity estimates with the \cite{Netzer2009} calibration. However, the estimates start to diverge at higher metallicities with high ionization parameter.  The low-metallicity offset in log($\mathrm{L_{bol,N09} / \Lacc}$) is higher for \PLalpha=-1.2, at $\sim1.0$, and lower at steeper \PLalpha, at $\sim0.64$ for \PLalpha=-2.0.  This is a parameter we cannot directly constrain from the fits, and so the absolute value for the bolometric luminosity will remain uncertain here. In the context of our results for SMACS S06355, the high derived NLR metallicity, log(\ZAGN/\Zsun) $= \newchangemath{0.12^{+0.14}_{-0.14}}$, couples it to a grid space corresponding to lower discrepancy between \cite{Netzer2009} bolometric luminosity estimates and the derived \Lacc\ from \beagleagn.

To investigate further the deviations from the \cite{Netzer2009} calibration, we reproduced their fig. 3 in \newchange{Figure}~\ref{fig:NetzerComparison_2}, which shows the logarithm of the ratio of a given emission line luminosity (either \Hbeta\ or \OIIIfi) against the logarithmic ratio between \OIIIfi\ and \Hbeta\ (a proxy for the ionization parameter).  Here we show the distribution of the models with $\PLalpha=-1.7$, for a range of metallicities and ionization parameters (see caption for details) as well as the measured value regions for a sample of Type-I AGN originally taken from \cite{Netzer2007}, though plotted in this form for the first time in \cite{Netzer2009}. We see that the models overlap with the observed Type-I measurements for the higher metallicities, but at lower metallicities they sit above the locally measured values (there is greater overlap for the steeper $\PLalpha=-2.0$). While $\log(\OIIIfi/\Hbeta)$ may show increasing ionization parameter at given metallicity, changes to metallicity with fixed ionization parameter has a more complicated form in this scheme.  We see that  $\log(\OIIIfi/\Hbeta)$ increases with decreasing \ZAGN\ from the highest metallicities, to a maximum value at which point it turns over and shows a decreasing trend with decreasing \ZAGN.  This demonstrates that a calibration based on \Hbeta\ alone, or including the ratio between \OIIIfi\ and \Hbeta, is insufficient to account for the complicated mapping of physical parameters on to these observables. Care is therefore required when applying these calibrations derived at low redshift to higher redshift samples that may well have much lower metallicities. In the case of SMACS S06355, the positions of the derived $\OIIIfi\mathrm{^{NLR}}$ and $\Hbeta\mathrm{^{NLR}}$ based points within Fig. \ref{fig:NetzerComparison_2} are in close proximity to the \ticks{elbow} of the \cite{Feltre2016} model grid tracks, an area sensitive to mis-mapping between the bolometric luminosity calibration of \cite{Netzer2009} and the \cite{Feltre2016} models.

Bolometric luminosities derived with \beagleagn\ will be able to self-consistently take account of the complicated response of emission line luminosities with respect to e.g. \logUsAGN\ and \ZAGN.  However, they will be highly dependent on the constraints placed on $\PLalpha$ and will be fundamentally degenerate with the chosen covering fraction of the NLR (10\% in this case).  Taking these considerations into account, we find the most consistent comparison to other AGN found at high redshifts comes with comparing to derivations of the bolometric luminosity derived from the same set of NLR models, as was performed by \cite{Scholtz2023}. In that work, the 42 Type-II AGN candidates (1 $\lesssim$ $z$ $\lesssim$ 10) were identified through use of various optical and UV emission line diagnostics, in concert with demarcation lines set in conjunction with the \cite{Feltre2016} and \cite{Gutkin2016} NLR and star-forming models, respectively. Here, the bolometric luminosity estimates of these objects were calculated using new calibrations from Hirschmann et al. (in prep), which uses the \cite{Feltre2016} models with the same \PLalpha\ slope. In this comparison, we note our derived accretion disc luminosity of log(\Lacc / $\mathrm{ergs^{-1}}$) = $45.19^{+0.12}_{-0.11}$ is higher than the range of bolometric luminosities presented in \citet[41.5 $\lesssim$ log($\mathrm{L_{bol} / ergs^{-1}}$) $\lesssim$ 44.5]{Scholtz2023}. This could suggest SMACS S06355 may be a particularly luminous AGN relative to other Type-II sources, especially given that our estimate disentangles the \HII\ and NLR contribution to the emission line flux, whereas the estimates in \cite{Scholtz2023} assume the flux is predominantly coming from the NLR.

\subsection{\NeIII\ emission line uncertainties}
\label{subsec:NeIII}

\begin{figure*}
\includegraphics[trim={1.45cm 0.5cm 0.25cm 0.25}, width=0.75\textwidth]{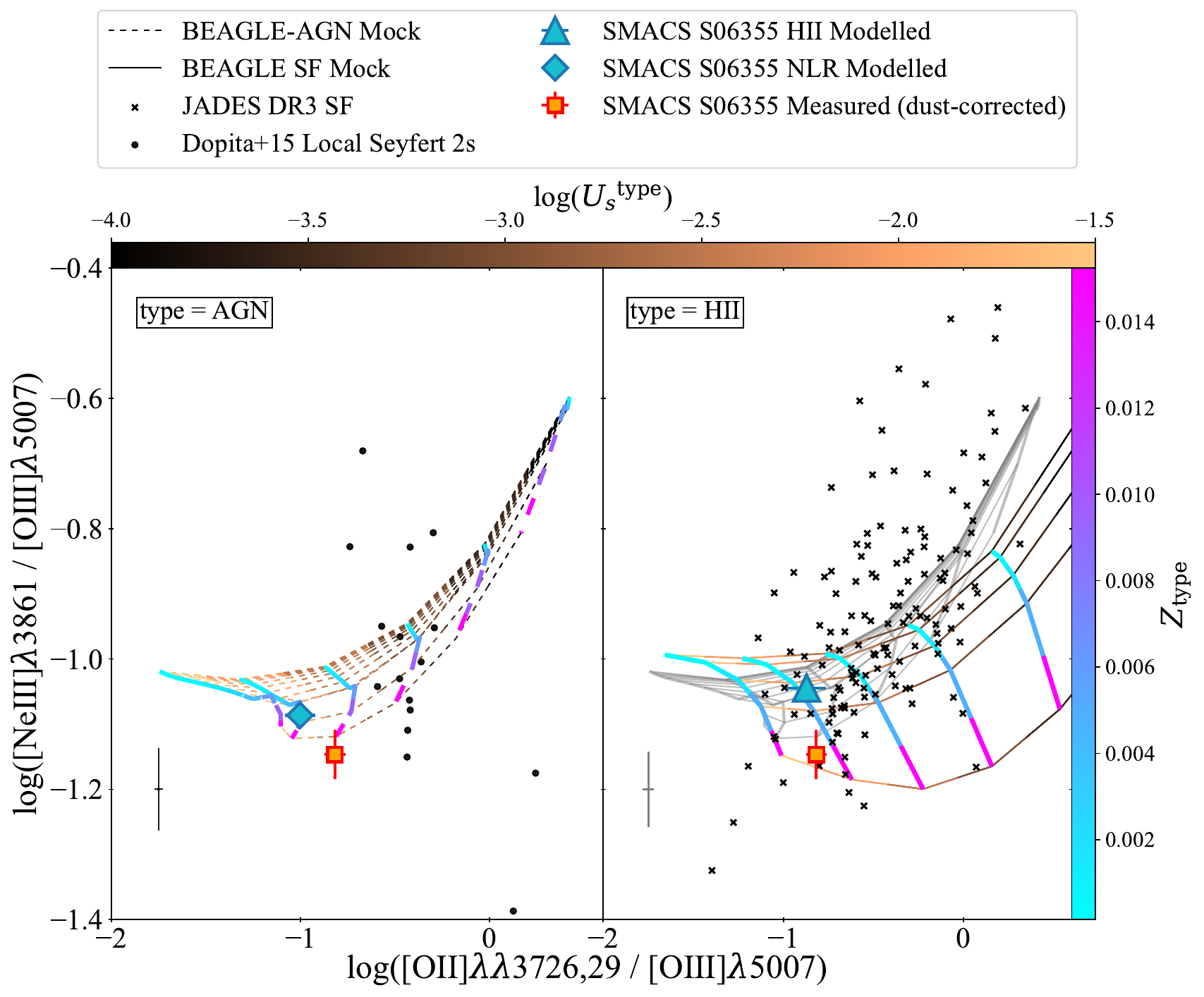}
\caption{\newchange{An emission line ratio plot, with $\mathrm{log(\NeIII/\OIIIfi)}$ and $\mathrm{log(\OII/\OIIIfi)}$ on the vertical and horizontal axes respectively, split according to whether the emission lines are dominated by star-forming (right panel) or narrow-line region (left panel) contribution.  The left panel shows the \protect\cite{Feltre2016} NLR model coverage for \PLalpha=-1.7, \xidAGN=0.3, \nHAGN=1000 cm$^{-3}$ as the dashed grid.  The narrow lines are colour-coded by \logUsAGN, as indicated in the upper colour-bar, and the thicker lines are colour-coded by \ZAGN\ as indicated by the colour-bar on the right.  Circular markers represent line ratios measured from the sample of \protect\cite{Dopita2015} $z<0.02$ NLR AGN from the \textit{Siding Spring Southern Seyfert Spectroscopic Snapshot Survey}, with average errors indicated as the black marker in the bottom left.  The square plus error bars (both panels) shows the line ratios measured for SMACS S06355 (corrected for dust using the measured \Hgamma/\Hbeta\ ratio, and assuming an intrinsic, Case B ratio of 0.468).  The diamond shows the modelled line rations from the fiducial} \beagleagn\ \newchange{fit, after separating out the NLR component \textit{only}.  The right panel shows the model coverage from the \protect\cite{Gutkin2016} \HII\ model coverage for \xid=0.3 and \nH=100$^{-3}$ as the grid (same colour-coding as in the left panel).  The light grey lines indicate the coverage of the NLR grid from the left panel for ease of comparison.  Objects from the most recent JADES data release \protect\citep[DR3,][]{D'Eugenio2024} are over-plotted for comparison as black crosses, with average errors indicated as the grey marker in the bottom left.  These line measurements have been corrected for dust attenuation using the \Halpha/\Hbeta\ Balmer decrement, assuming Case B recombination. Additionally, any JADES objects identified as AGN candidates from \protect\cite{Scholtz2023} have been removed.  The triangle shows modelled line ratios for the fidicual} \beagleagn\ \newchange{fit, after separating out the \HII\ component.}}
\label{fig:NeIII_EL_Plot}
\end{figure*}

As mentioned in Section~\ref{subsec:data}, we refrain from including \NeIII\ \newchange{\citep[also measured by][]{Curti2023a}} in the SED modelling.  Inclusion of this line leads to significantly higher NLR metallicities (log($\ZAGN/\Zsun) = \newchangemath{0.23^{+0.05}_{-0.05}}$).  However, it has been noted in other works that the \NeIII\ emission of high-redshift galaxies follow an unexpected behaviour, such as \citet[fig. 6]{Shapley2024}, who find high redshift galaxies with \NeIII/\OII\ ratios extending to higher values compared to local samples. This includes part of their sample of $1.4<z<7$ star-forming galaxies and the $5.5\lesssim z \lesssim9.5$ spectral stacks of \cite{Roberts-Borsani2024}. We show in Figure \ref{fig:NeIII_EL_Plot} \newchange{(right panel)} a comparison of star forming galaxies taken from the JADES third data release \cite{D'Eugenio2024} to the model coverage of $\log(\NeIII/\OIIIfi)$ vs. $\log(\OII/\OIIIfi)$.  We see that at all values of \OII/\OIIIfi, there is a spread of measured \NeIII/\OIIIfi\ to higher values than covered by the models.  The persistence of this trend to high \OII/\OIIIfi\ (or low \logUs) dis-favours an explanation based on the difference in ionising source, since we expect a higher \NeIII/\OIIIfi\ ratio to be driven by harder ionising field which would in turn decrease \OII/\OIIIfi.  It is therefore more likely that the difference is driven by \newchange{variation in gas properties not covered by the model grid used in the fiducial} \beagle\ \newchange{fit (e.g. }a difference in abundance of Neon compared to Oxygen in the gas phase).  We do not see a similar mis-match between model coverage and observations in other key ratios (e.g. \OIIIfi/\Hbeta) and so infer that the issue is with an under-estimated \NeIII\ flux in the models.  A consequence of this under-estimation within the models would be to attribute more of the \NeIII\ flux to the NLR component, hence driving up the derived metallicity \newchange{in the absence of varied Ne/O abundance in the \cite{Feltre2016} models, which assumes a solar value $\mathrm{log(Ne/O) = -0.84}$}.

\newchange{We also show the measured line ratios for SMACS S06355 in Fig.~\ref{fig:NeIII_EL_Plot} in comparison to the \HII\ region model grid (right panel) and NLR model grid (left panel). These model grids are presented as nets color-coded by both ionisation parameter \logUs\ and metallicity \Z, where our models assume a solar metallicity of $\Zsun\ = 0.01524$ \citep{Gutkin2016, Feltre2016}. Despite the JADES star-forming galaxies extending above the model grid, SMACS S06355 is in a region of low \NeIII/\OIIIfi\ and has \HII\ region model coverage.  However, we caution that we expect there to be an \HII\ region \textit{and} NLR contribution to all the emission lines present in these line ratios, and so a direct comparison of the measured line ratios to the model grids is unhelpful.  We therefore show the modelled line rations for the \HII\ region and NLR contributions for the fiducial} \beagleagn\ \newchange{fit (which did not fit explicitly to \NeIII\ for the reasons outlined previously) separately in the corresponding figure panels. However, we are limited by the coverage of the underlying models and cannot determine whether the true \NeIII/\OIIIfi\ ratio of \HII\ region emission in this galaxy is consistent with the models.  Nonetheless, we find that the ratio predicted from pure \HII\ emission is significantly higher than the measured value.}

\newchange{For interest, we also display measured line ratios for a local sample of NLR AGN taken from the \textit{Siding Spring Southern Seyfert Spectroscopic Snapshot Survey} \citep{Dopita2015} in the left panel of Fig.~\ref{fig:NeIII_EL_Plot}.  We note that the ratios extend to similarly high \NeIII/\OIIIfi\ values as the high-redshift star-forming galaxies shown in the right panel.  There is also a set of low \NeIII/\OIIIfi\ ratios at high \OII/\OIIIfi\ ratios that are not displayed by the star-forming galaxies, and that cannot be explained by the current NLR modelling.}

If this galaxy does have a higher Neon-to-Oxygen abundance in the gas-phase than local galaxies, then our NLR metallicity constraints would also likely be over-estimated.  We caution, therefore, that since our main anchor to the NLR contribution is given by \NeIV, our NLR constraints are tied inextricably to the relative Neon abundance in the NLR itself.  Our high derived NLR metallicity estimate may instead be a marker of a higher Neon-to-Oxygen abundance in this galaxy than present in our models.

\subsection{Direct Neon and Oxygen abundance calculations}
\label{subsec:direct_abundances}

\newchange{Given the set of modelled emission lines, we  can calculate `direct' Oxygen and Neon abundances using calibrations appropriate for NLR AGN, though caution that these methods do not consider possible \HII\ region contribution to the lines.}

\newchange{We first calculate the Oxygen abundance following \cite{Dors2020}.  Essentially the electron temperature is estimated for the region containing the O$^{2+}$ ions ($t_3$) by using the measured \OIIIauroral\ and \OIIIfi\ fluxes via their equation 1.  Since we do not have a flux measurement for an auroral line for the O$^{+}$ ion, we then must estimate the electron temperature in the low-ionization region, and \cite{Dors2020} provided the first NLR appropriate calibration for this temperature estimate based on $t_3$ (their equation 2).  Then the O$^{2+}$/H$^{+}$ and O$^{+}$/H$^{+}$ ionic abundances are calculated employing $t_3$ and $t_2$ (their equations 3 and 4).  However, it is non-trival to calculate the ionization correction factor (ICF) required to account for higher ionization states of Oxygen.  For an ICF of 1, we find 12+log(O/H) = $8.31 \pm 0.38$ but we know from the presence of the \NeIV\ line that the ICF(O) should be significant for this galaxy. In the absence of other constraints, we resort to the maximum measured ICF(O) from the sample in \cite{Dors2020} of 1.39 as a baseline, finding 12+log(O/H) = $8.45 \pm 0.38$.  If we compare these two estimates to the} \beagle\newchange{-derived value, 12+log(O/H)$^{\mathrm{NLR}}$ $=8.86^{+0.14}_{-0.16}$ (akin to using ICF(O) = 1.40\footnote{\newchange{This was calculated by estimating the ${\rm O}^{2+}/{\rm H}^++{\rm O}^+/{\rm H}^+$ abundance from the NLR contributions to \OIIIauroral\ and \OIIIfi\ line fluxes following the \citet{Dors2020} method, and comparing to the} \beagle\newchange{-derived total Oxygen abundance, 12+log(O/H)$^{\mathrm{NLR}}$}.}), we find best agreement (at $\sim 1 \sigma$) with the ${\rm ICF(O)}=1.39$ based calculation.}

\newchange{To estimate the `direct' Ne/H abundance we use the method of \citet{Armah2021}.  They investigate the use of $t_3$, calculated via the \OIIIauroral\ and \OIIIfi\ lines, to estimate the Ne$^{2+}$/H$^{+}$ abundance. Using cloudy modelling, they find a non-linear ratio between the average electron temperature in the region hosting the Ne$^{2+}$ ion, \teNeIII, and $t_3$ for the region hosting O$^{2+}$, where \teNeIII\ is lower than $t_3$.  This is because the Ne$^{2+}$ ion inhabits a much more extended region than O$^{2+}$ (see their figure 5),  and we verified that the \cite{Feltre2016} models showed the same behaviour.  We use their Ne$^{2+}$ ICF calibrated from measured lines in the infrared (their equation 34) that accounts for Ne$^+$ ions.  We note, however, that higher ionisation states are not corrected for, suggesting that this method will provide a lower limit to the Neon abundance in this galaxy.  Using $t_3$ to estimate the electron temperature of the high ionization zone, we obtain 12+log(Ne/H) = $7.84 \pm 0.65$, while \teNeIII\ gives 12+log(Ne/H) = $8.15 \pm 0.51$.}

\begin{table}
\begin{center}
\caption{\newchange{Direct estimations of $\mathrm{log(Ne/O)}$ abundance based on the methods of \protect\cite{Armah2021}, explained further in text, using different ICFs (1,  1.39, 1.40) and electron temperatures ($t_{3}$ and $t_{e}\mathrm{(NeIII)}$). The solar value is taken to be $\mathrm{log(Ne/O) = -0.84}$, calculated from table 1 of \protect\cite{Gutkin2016}.}}
\begin{tabular}{ |c|c|c| } 
 \hline
 ICF(O) & $\mathrm{t_{3}}$ & \teNeIII\\
 \hline
 1 & $-0.47 \pm 0.04$ & $-0.16 \pm 0.03$ \\ 
  1.39 & $-0.62 \pm 0.04$ & $-0.31 \pm 0.03$ \\ 
 1.40 & $-0.60^{+0.14}_{-0.12}$ & $-0.29^{+0.08}_{-0.07}$ \\ 

 \hline
\end{tabular}
\label{tab:neon_table}
\end{center}
\end{table}

\newchange{We combine these O/H and Ne/H abundances to estimate Ne/O, and provide the range of estimates for different ICF(O) values and electron temperature estimates experienced by Ne$^{2+}$ in Table~\ref{tab:neon_table}.  The large range of values highlights the uncertainty in the Ne/O estimate.  All values indicate super-solar Ne/O abundances, which in AGN are not necessarily unusual; for instance, some of the AGN sources studied within \cite{Armah2021} also reached super-solar 
Ne/O.}

\newchange{While \citet{Armah2021} suggest using \teNeIII, we note that flattening of the relation between $t_3$ and \teNeIII\ indicates that it is not optimal to use emission lines of Oxygen to estimate the electron temperature relevant for Ne$^{2+}$ ions, contributing considerable uncertainty to our results.  Moreover, we are unable to correct for ionisation states higher than Ne$^{2+}$, meaning that pairing the derived Ne/H with our preferred total `direct' oxygen abundance estimate (akin to ICF(O) = 1.40) may lead to an underestimated Ne/O ratio.} 

\newchange{\citet{Arellano-Cordova+22} also provided an estimate of the Ne/O abundance for this galaxy, but they assumed it was star-forming.  They estimated $\mathrm{log(Ne/O)}=-0.64 \pm 0.17$, which is consistent with our estimates despite the different underlying assumptions of ionizing source.}

\subsection{Dependence of results on modelling of NLR dust}
\label{subsec:nlr_dust}

In modelling the dust attenuation experienced by \HII\ regions and the NLR, we took the simplest approach that avoided adding an additional free parameter by allowing the NLR to only be attenuated by the dust in the diffuse ISM (see description in Section \ref{subsec:beaglefitting}).  However, Fig.~\ref{triangleplot} shows that there is a clear degeneracy between the $V$-band optical depth, \tauV\ and the accretion disc luminosity, \Lacc.  The photometry should provide independent constraints on the dust attenuation experienced by stars since it covers the UV slope.  However, our simple model tied the attenuation of the NLR to that of the stars.  

We tested how this introduced degeneracy might be affecting the other parameter constraints by setting up a fit with separate dust attenuation curve for the NLR, parameterised as:
\begin{equation}
    \hat{\tau}_{\lambda}^\textsc{NLR} = \tauVnlr\left(\frac{\lambda}{0.55\mu\textrm{m}}\right)^{-n_V^\textsc{NLR}}
\end{equation}  

This required introducing a further free parameter, \tauVnlr, which is the $V$-band optical depth of the dust attenuation of the NLR.  When setting $n_V^{\textsc{NLR}}$ we chose two limiting values, a slope similar to the stellar birth clouds ($1.3$) and one similar to the ISM ($0.7$) and compare the results.  The fits show very poor constraints on the extra free parameter \tauVnlr\ in both scenarios while the constraints on \tauV\ remain virtually unchanged, showing that the dust constraints have indeed been driven by the star-forming component via the photometry.  Most parameters show no statistically significant changes with respect to the fiducial fit, though a possible trend of decreasing \Zhii\ and \logUs\ are seen with increasing NLR dust slope.  The uncertainties on \Lacc\ also increase significantly for dust slope of 0.7 ($\log(\Lacc / \mathrm{ergs^{-1}})=45.48^{+0.34}_{-0.35}$), though are similar for slope of 1.3 ($45.44^{+0.15}_{-0.15}$), showing that the bolometric luminosity constraints are sensitive to the modelling of NLR dust.

\subsection{Dependence of results on aperture corrections applied to emission line fluxes}
\label{subsec:aperture_corrections}

As mentioned in Section \ref{subsec:data}, the method used to correct the measured emission line fluxes to `total equivalent' fluxes implicitly assumes that the morphology of the line-emitting gas follows that of the continuum.  This is a simplification that may be appropriate for small galaxies that have a reasonable fraction of their flux captured by the MSA in order to estimate e.g. total \Halpha-based SFRs.  However, it is clearly inappropriate when there are contributions to the emission line fluxes from different morphological extents; in this case, extended star formation as well as a NLR contribution from a point-like nuclear region.  It will likely result in over-estimating the fractional AGN contribution to various emission lines as well as the AGN bolometric luminosity.  Though it may be possible to dis-entangle the nuclear and extended regions from the continuum light with image decomposition using tools such as e.g. \textsc{forcepho} (Johnson et al., in prep), we lack the information of how the emission line fluxes should be attributed to the two components without integral-field spectroscopy.  

We test the likely effects of the over-correction of NLR components by fitting using the un-corrected \NeIV\ flux, but leaving unchanged the photometry and other line fluxes that were used in the fiducial fitting.  We note that the NLR contribution to other emission line fluxes will likely still be over-estimated, but this test should be sufficient to provide an indication of how this might affect our results.  We find minimal change in dust attenuation ($\newchangemath{\tauV=0.63^{+0.12}_{-0.12}}$) and \HII\ region metallicity ($\newchangemath{\log(\Zhii/\Zsun)=-0.64^{+0.17}_{-0.15}}$), which are comparable to the fiducial within the uncertainties.  There are much poorer constraints on the NLR metallicity, $\newchangemath{\log (\ZAGN/\Zsun)=-0.04^{+0.27}_{-0.39}}$, and lower accretion luminosity at the $\sim2\sigma$ level, $\log(\Lacc / \mathrm{ergs^{-1}})=44.93^{+0.07}_{-0.08}$, but comparable SFR, $\log(\sfrInLog/\Msun\yr^{-1})=1.70^{+0.04}_{-0.05}$.  This indicates that the NLR metallicity is indeed driven by the \NeIV\ flux estimate and is therefore dependent on the chosen correction to total.  

In future work we will improve the modelling of the fractional contribution of the NLR and \HII\ region components. Though this is beyond the scope of this paper, future work may include fitting the aperture corrections themselves, such that the total photometry-sensitive correction (like that described in Section \ref{subsec:data}) can be set as an upper prior limit. This may mitigate some difficulty in \ticks{choosing} which emission lines are to be considered nuclear in distribution (and therefore correction), which would allow flexibility especially for middle range ionising potential lines.

\section{Summary and Conclusions}
\label{section:summary}

In this work we presented a characterisation of the Type-II AGN candidate SMACS S06355 at $z$ $\approx$ 7.6643, employing \beagleagn\ \citep{Vidal-Garcia2024}, the extension of \beagle\ \citep{Chevallard2016} which self-consistently incorporates a prescription for the AGN narrow-line region \citep{Feltre2016} into the \HII\ region modelling. Using this, we fitted to the \NeIV, \OII, \Hdelta, \Hgamma, \OIIIft, \Hbeta, \OIIIfn, and \OIIIfi\ emission line fluxes, themselves originating from the G235M/F170LP and G395M/F290LP grating/filter configurations of ERO NIRSpec MSA data from the SMACS 7327 cluster field, and corrected to their `total equivalent' values. We also used the following NIRCam filters in our fitting: F090W, F150W, F200W, F277W, F356W, and F444W. With the \NeIV\ emission line requiring ionising sources other than star formation, we were able to constrain AGN related parameters by including it in our fitting, namely log(\ZAGN/\Zsun), log(\Lacc), and log(\UsAGN). From our characterisation of this galaxy, the key findings and interpretations are:

\begin{itemize}
    \item Our independent measurement of the \NeIV\ flux, extracting from the G235M/F170LP NIRSpec MSA spectrum, yielded an observed value of $\mathrm{4.39 \pm 0.42 \times 10^{-19} erg s^{-1} cm^{-2}}$, uncorrecting for magnification. The extraction method was verified as being consistent with those associated with the \OII, \Hdelta, \Hgamma, \OIIIft, \Hbeta, \OIIIfn, and \OIIIfi\ fluxes from \cite{Curti2023a}, and so we could then convert all fluxes together to forms consistent with the photometry based morphology.
    \item In comparison to previous studies of the same source, our characterisation presents a marginally higher SFR (log(\sfrInLog / $\mathrm{\Msun\ yr^{-1}}$) = $1.70^{+0.10}_{-0.09}$), driven by our higher derived \tauV, and perhaps also linking to the use of `total equivalent' fluxes as opposed to ones which assume point source corrections. Intuitively, this SFR decreases upon switching to an IMF with 300 $\mathrm{M_{\odot}}$ as an upper mass limit. Our fiducial gas phase metallicity is also approximately $\newchangemath{0.41}$ dex lower than that presented in \cite{Curti2023a}. We attribute this to a difference in methodology, namely the direct T$_{\mathrm{e}}$ method used in \cite{Curti2023a} attributing total flux to star formation, whereas our derived value being made through comparisons to photoionization models within the SED fitting, whilst considering both SF and AGN components. 
    \item The derived gas-phase metallicity for the NLR component, 12 + log(O/H)$\mathrm{^{NLR}} = \newchangemath{8.86^{+0.14}_{-0.16}}$, is unconstrained in our fitting. The 95\% probability region of the metallicity posterior extends to subsolar values, with the 68\% probability region gravitating towards solar values. These large uncertainties generally arise from there being just a single high-ionization line (uncontaminated by star formation) available to constrain NLR related parameters in this case. Additionally, the high derived metallicity is likely a result of our models' current treatment of Neon abundances, in comparison to the emerging high \NeIII/\oii\ ratios seen in high redshift galaxies. 
    \item Our fit to SMACS S06355 includes a significant AGN contribution to lines such as \OIIIfi\ and \Hbeta; this is accompanied by a moderate accretion disc luminosity log(\Lacc / $\mathrm{ergs^{-1}}$) = $45.19^{+0.12}_{-0.11}$, which we note is slightly higher than comparable Type-II AGN across a range of cosmic time, namely those in \cite{Scholtz2023}. Our bolometric luminosity estimate calculated from the calibration of \cite{Netzer2009}, $\mathrm{log(L_{bol} / ergs^{-1}) = 45.67 \pm 0.33}$, is either smaller than or comparable to other Type-I AGN studied throughout cosmic time (\citealt{Ubler2023, Maiolino2023b}).
    \item Whilst we present our characterisation of SMACS S06355 as including an AGN component, we simultaneously recognize the notion that high-ionization emission lines (such as \NeIV) may also originate from alternate sources to AGN. Given the emission line we are using to constrain AGN related parameters (\NeIV) \newchange{requires the presence of high-energy photons} ($\sim$ 63 eV), we recognize there may be typically less numerous alternative sources of ionization, though we do not completely rule out alternative ionising sources like X-ray binaries at this time.
    \item \newchange{We calculate `direct' estimates for Neon and Oxygen abundances using AGN-appropriate calibrations \citep{Dors2020, Armah2021}. Whilst demonstrating the large uncertainties involved with poorly constrained ionization correction factors and electron temperatures, our estimates potentially indicate a super-solar Ne/O abundance in SMACS S06355.} 
    \item Through additional testing, we show NLR parameters like accretion disc luminosity \Lacc\ are sensitive to the modelling of the NLR dust. This parameter, along with others such as dust attenuation optical depth \tauV\ and NLR metallicity \ZAGN\ are also sensitive to the aperture corrections performed, and hence the underlying physical assumptions. This highlights the importance in considering different modelling approaches, especially when applying these to early epochs.
\end{itemize}

This work has presented an instance of applying \beagleagn\ to a high redshift Type-II AGN candidate; despite there being only one line free from SF contamination to constrain NLR related parameters within our SED fitting, we have been able to derive some constraints on accretion disc luminosity log($\mathrm{\Lacc}$), NLR metallicity log(\ZAGN/\Zsun), and NLR ionization parameter log(\UsAGN). With new observations of obscured AGN enabled by JWST, it will be possible to apply \beagleagn\ in a similar manner, and characterise a larger number of Type-II AGN candidates; for example, those presented in \cite{Scholtz2023} (Silcock et al. in prep). Such objects with multiple high-ionization lines will also present opportunities to derive tighter constraints on both AGN contribution to galaxy SEDs and derived parameters. 
This work has highlighted some necessary improvements to the modelling and treatment of observables. Specifically: the NLR dust model; varying elemental abundances and uncertainties with spatial extent of emission lines if including photometry. These efforts will be able to contribute to our understanding of AGN demographics across cosmic time, including the comparisons to their hosts, general evolution and other items of discussion such as contributions to the Epoch of Reionization.

\section*{Acknowledgments}
\newchange{The authors are very grateful for correspondence with Mirko Curti and Sandro Tacchella, and thank the referee for their valuable and constructive comments, which have improved this work}.  MSS acknowledges support from an
STFC PhD studentship (grant ST/V506709/1). ECL acknowledges support of an STFC Webb Fellowship (ST/W001438/1). DJBS acknowledges support from the UK Science and Technology Facilities Council (STFC) under grants ST/V000624/1 and ST/Y001028/1. AJB, JC and IEBW acknowledge funding from the ERC Advanced Grant 789056 “FirstGalaxies” (under the European Union’s Horizon 2020 research and innovation programme). AVG acknowledges support from the Spanish grant PID2022-138560NB-I00, funded by MCIN/AEI/10.13039/501100011033/FEDER, EU. MH and AP acknowledge financial support from the Swiss National Science Foundation via the PRIMA grant “From cosmic dawn to high noon: the role of BHs for young galaxies” (PROOP2$\_$193577). AF acknowledges the support from project "VLT-MOONS" CRAM 1.05.03.07, INAF Large Grant 2022 "The metal circle: a new sharp view of the baryon cycle up to Cosmic Dawn with the latest generation IFU facilities" and INAF Large Grant 2022 "Dual and binary SMBH in the multi-messenger era". SC acknowledges support by European Union’s HE ERC Starting Grant No. 101040227 - WINGS. The authors acknowledge support from the European Research Council (ERC) via an Advanced Grant under grant agreement no. 321323-NEOGAL.

\section*{Data Availability}
The data underlying this article are available in the article and/or upon request.

\bibliographystyle{mnras}
\bibliography{bibliography.bib}

\label{lastpage}

\end{document}